\documentclass[%
 reprint,
 amsmath,amssymb,
 aps,
]{revtex4-1}

\usepackage{graphicx}% Include figure files
\usepackage{dcolumn}% Align table columns on decimal point
\usepackage{bm}% bold math
\usepackage{hyperref}% add hypertext capabilities
\usepackage{amsthm}
\usepackage[caption=false]{subfig}

\begin{document}

\title{Voltage-Controlled Topological-Spin Switch for Ultra-Low-Energy Computing--Performance Modeling and Benchmarking}

\author{Shaloo Rakheja}
 \email{sr3936@nyu.edu}
 \affiliation{Department of Electrical and Computer Engineering, New York University, Brooklyn, NY 11201, USA
 }
\author{Michael E. Flatt\`{e}}%
 \email{michael flatte@mailaps.org}
\affiliation{Department of Physics and Astronomy and Optical Science and Technology Center, University of Iowa, Iowa City, Iowa 52242, USA
}

\author{Andrew D. Kent}
\email{andy.kent@nyu.edu}
\affiliation{Center for Quantum Phenomena, Department of Physics, New York University, New York, NY 10003, USA
}


\begin{abstract}
A voltage-controlled topological-spin switch (vTOPSS) that uses a hybrid topological insulator-magnetic insulator multiferroic is presented that can implement Boolean logic operations with sub-10 aJ energy-per-bit and energy-delay product on the order of $10^{-27}$ Js.
%This paper presents a voltage-controlled topological-spin switch (vTOPSS) that uses a hybrid topological insulator-magnetic insulator multiferroic to implement Boolean logic operations with sub-10 aJ energy-per-bit and energy-delay product on the order of $10^{-27}$ Js.
The device uses a topological insulator (TI), which has the highest efficiency of conversion of electric field to spin torque yet observed at room temperature, and a low-moment magnetic insulator (MI) that can respond rapidly to a given spin torque. 
%Due to its insulating nature, the MI layer does not shunt the electric current, which is restricted to flow on the surface of the TI layer.
%The magnetoelectric transduction efficiency of vTOPSS surpasses that of existing magnetoelectric structures by orders of magnitude and allows for voltages as low as few 10's of mV for reliable operation. 
We present the theory of operation of vTOPSS, develop analytic models of its performance metrics, elucidate performance scaling with dimensions and voltage, and benchmark vTOPSS against existing spin-based and CMOS devices. 
Compared to existing spin-based devices, such as all-spin logic and charge-spin logic, vTOPSS offers 100$\times$ lower energy dissipation and (40-100)$\times$ lower energy-delay product. 
%At the same time, the performance--energy, delay, and energy-delay product--of vTOPSS is comparable to that of the magnetoelectric spin-orbit logic.
%While the delay of vTOPSS is comparable to that of the 
%magnetoelectric spin-orbit logic, its energy dissipation is (2-5)$\times$ lower. 
With experimental advances and improved material properties, we show that the energy-delay product of vTOPSS can be lowered to $10^{-29}$ Js, competitive against existing CMOS technology. Finally, we establish that interconnect issues that dominate the performance in CMOS logic are relatively less significant for vTOPSS, implying that highly resistive materials can indeed be used to 
interconnect vTOPSS devices.
\end{abstract}

\maketitle

\vspace{-10pt}
\section{Introduction}
\vspace{-10pt}
Spin-based logic and memory devices use nanomagnets as digital spin capacitors to store and manipulate information~\cite{chappert2007emergence}. 
Typically, spin-polarized electrical currents or magnetic fields are used to control the magnetization vector of nanomagnets while reading and writing information~\cite{locatelli2014spin}. Compared to their charge-based counterparts, spin-based devices offer non-volatility of information and superior logical efficiency, i.e. fewer devices to implement a given Boolean function~\cite{kani2014pipeline}. However, the majority of spin-based devices suffer from high energy dissipation 
resulting from a large electric current density on the order of $10^6$ A/cm$^2$ required to reorient the magnetization vector~\cite{huai2008spin, liu2012spin}. Such large current densities not only lead to excessive Joule heating in the device, but could cause electromigration issues in metallic interconnects~\cite{katine2008device}. At the same time, reversal of metallic ferromagnetic bodies using anti-damping spin-transfer torque (STT) proceeds on a timescale on the order of 100's of picoseconds to a few nanoseconds~\cite{nikonov2010strategies}. As such, existing spin-based devices have an energy-delay product that is (1,000-10,000)$\times$ larger than that of their CMOS counterparts~\cite{nikonov2015benchmarking}. 
% increasing the enersgy-delay product of spin-based devices 

To harness the full potential of spintronics technology, it is imperative to develop methods for energy-efficient and fast manipulation of the magnetic order parameter. Actuation methods, such as
%based on 
voltage control of magnetic anisotropy and coercivity, use of magnetoelectric and exchange coupling in 
multiferroic/ferromagnetic heterostructures, and 
charge carrier density mediated ferromagnetism control, have been investigated~\cite{chu2008electric, duan2008surface}. 
Yet, these effects are generally weak at room temperature, which limits their practical use. 
For example, full 180$^\circ$ reversal of a ferromagnet via the magnetoelectric effect requires the assistance of electric currents or magnetic fields or can be accomplished using the resonant pulsed switching mode that requires precise pulse timing~\cite{wu2011giant, hu2015purely, peng2016fast, kani2017strain}. 
Magnetoelastic effects that are used to tune the magnetic properties of thin films via epitaxial strain or piezoelectric substrates are generally observed in small aspect ratio nanomagnets~\cite{tiercelin2011room, tiercelin2011magnetoelectric}. However, in high aspect ratio nanomagnets it is difficult to use strain effects to tune the magnetic properties.

A promising research direction is ``topological spintronics'' that has been driven by the demonstration of efficient room-temperature spin-charge conversion in heterostructures 
%in which 
%that interface 
with
a topological insulator (TI) 
%with
interfacing 
a ferromagnetic (FM) metal~\cite{mellnik2014spin}. The key property responsible for this advance is the combination of large spin-orbit coupling (SOC) strength and time-reversal symmetry that leads to the formation of helical Dirac surface states possessing an inherent spin-momentum locking~\cite{Burkov2010,Hasan2010, burkov2010spin, witczak2014correlated}. 
The distinctive feature of TIs is that even without carriers near the chemical potential in bulk, the spin Hall conductivity can be finite and significantly larger than that of heavy metals such as Pd, Pt, W~\cite{guo2008intrinsic, hoffmann2013spin, khang2017conductive}.

Here, we utilize electric fields across a TI 
resulting in a coherent transport of spins 
across the material to generate a spin torque on the magnetization of an adjacent magnetic insulator (MI) layer~\cite{lv2017unidirectional}.  
Unlike FM metal, there is no shunting of electric current in the MI layer and current is restricted to flow on the surface of the TI layer. Furthermore, the MI can induce a gap in the TI surface states, rendering the TI surface state insulating, which is (counterintuitively) beneficial to the device operation.
The spin-based device, {\bf{v}}oltage-controlled {\bf{top}}ological-{\bf{s}}pin {\bf{s}}witch (vTOPSS), decouples the elements of a magnetoelectric material~\cite{fiebig2005revival}, allowing us to simultaneously optimize the choice of both TI and MI materials, thereby enabling ultra-low-energy computing. One of the most important properties of the MI layer is its low damping~\cite{wu2013recent, heinrich2011spin} which is highly desirable in device applications where switching is realized through magnetization precession, as in vTOPSS.
The remainder of this paper is organized as follows. In Section II, the physics of operation of vTOPSS is presented. In Section III, analytic models of performance metrics of vTOPSS are presented followed by benchmarking results against existing spin- and charge-based devices in Section IV. 
In Section V, implementation of universal Boolean logic gates and the logical efficiency of vTOPSS resulting from its innate polymorphism 
%for large-scale circuit implementations 
are highlighted. Section V summarizes the key findings of this work while also offering an outlook on future research directions.

\vspace{-20pt}
\section{Physics of Operation}
\vspace{-10pt}
The evolution  of the wave functions of the  full bands of the TI, under the influence of an electric field, produces coherent transport of spins across the material~\cite{Sahin2015}, which can be used to efficiently manipulate the magnetization state of an adjacent magnetic layer~\cite{Flatte2017}. The charge Hall conductivity and bulk dissipative charge currents vanish or are small in the TI, but the spin Hall conductivity is finite and can be much larger than that of a large-SOC metal~\cite{Krempa2014}. A TI has the highest efficiency of conversion of electric field to spin torque yet observed at room temperature~\cite{Mellnik2014,Yabin2014}. 
Hybrid TI-MI structures decouple the constituent features  of a multiferroic  material, allowing independent optimization of both components of the response of magnetization to an electric field, i.e. generation of spin torque from the electric field and response of the magnetic moments to the spin torque.

The total Berry curvature of a full band measures the integrated correlation between spin and orbital degrees of freedom. For a so-called trivial insulator this correlation integrates to zero across the entire full band. Thus, if at one region of the Brillouin zone  the wave functions of the band have spin and orbit correlated preferentially parallel, there will be another region of the zone in which the wave functions are correlated preferentially antiparallel. An example is the valence band in a trivial direct-gap semiconductor, such as GaAs, which is of $p$-orbital character, for which the wave functions near the valence maximum are heavy hole states, with spin and orbit degrees of freedom parallel, whereas at  energies below the split-off energy the spin and orbit degrees of freedom are preferentially oriented antiparallel.

TIs differ from these trivial insulators in that this spin-orbit correlation does not integrate to zero. The spin-orbit correlation is described quantitatively by the Berry curvature of the band, and thus the electronic ground in a TI  possesses a nonzero integrated Berry curvature. The spin Hall conductivity in the clean static limit, evaluated as the linear response of the spin current to an electric field using the Kubo approach, depends directly on the Berry curvature~\cite{Guo2008}: 
\begin{equation}
\sigma_{yx} = \frac{e \hbar}{V}\sum_k f_{n {\bf k}} \Omega_{n{\bf k}}^z,
\end{equation}
 where $e$ is the elementary charge, $\hbar$ is the reduced Planck's constant, $V$ is the volume of the system, ${\bf k}$ is the crystal momentum, $n$ is a band index, and is $\Omega_{n{\bf k}}^z$ is the Berry curvature:
\begin{equation}
\Omega_{n{\bf k}}^z=2\sum_{n \neq n'} \mathrm{Im} \frac{\left<u_{n {\bf k}}|j_y^z|u_{n'{\bf k}}\right>\left<u_{n'{\bf k}}|v_x|u_{n{\bf k}}\right>}{(E_{n{\bf k}}-E_{n'{\bf k}})^2}.
\end{equation}
Here, the Fermi-Dirac function $f_{n {\bf k}}$ ensures that the sum is over filled states, corresponding to all the filled bands at zero temperature. 
The spin current and velocity operators, $\hat j_i^j$ and $\hat v_i$, are
\begin{align}{ \label{eq:spincurrent}
\hat j_i^j=\frac{\hbar}{4}(\hat v_i\sigma_j+\sigma_j\hat v_i), \qquad \hbar \hat v_i=\nabla_{k_i}\hat H,}
\end{align}
where $\sigma_j$ is the spin operator along direction $j$, and $\hat H$ is the Hamiltonian of the material. The current and velocity operators are evaluated between the  states with Bloch functions $u_{n\mathbf{k}}$ and $u_{n'\mathbf{k}}$, and energy $E_{n \mathbf{k}}$.

\begin{figure}[h!]
%\centering
\includegraphics[width=3.25in]{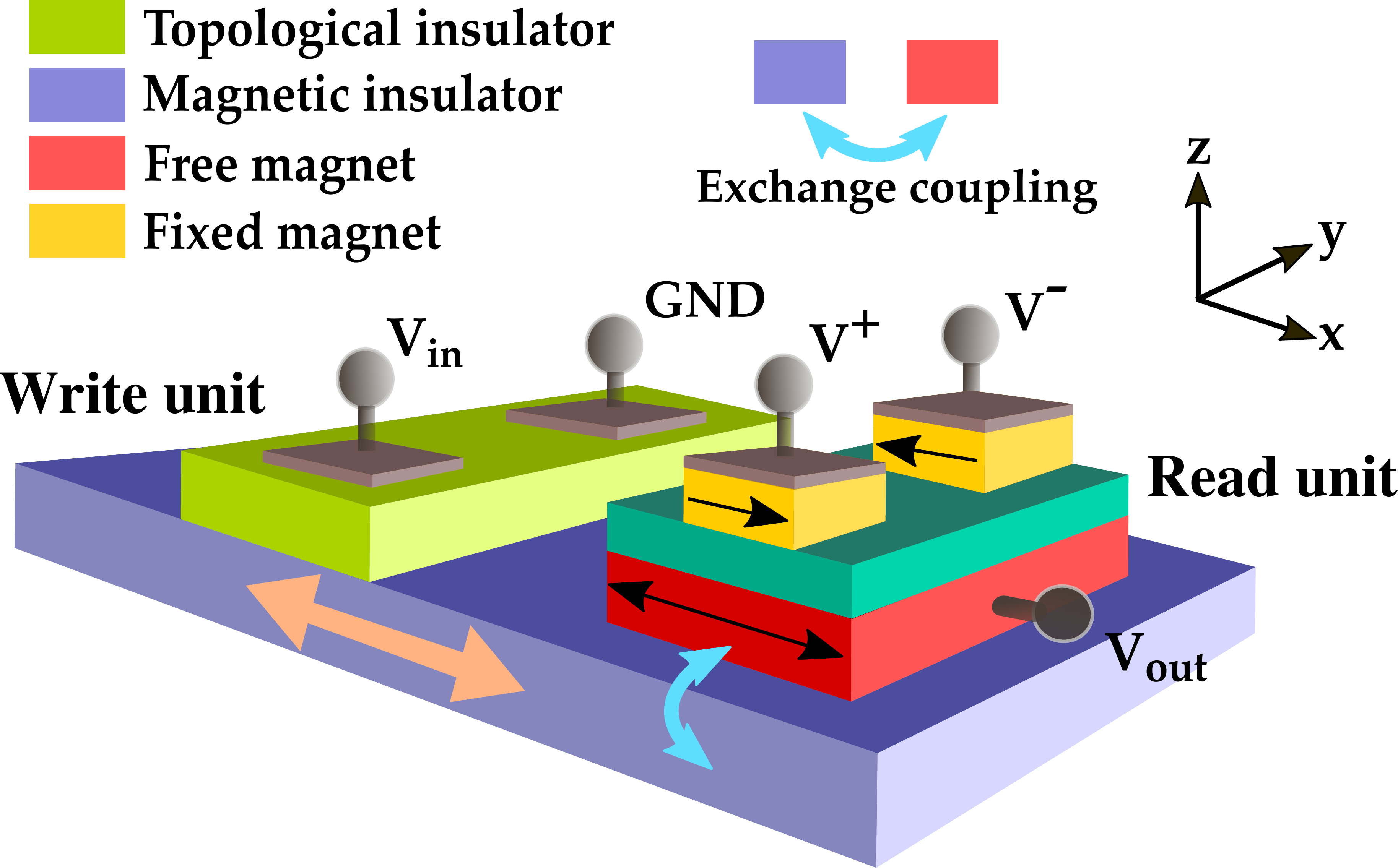} 
\caption{Copy/invert functions implemented using vTOPSS. In the write unit, input voltage signal applied across the TI layer creates spin accumulation at the interface of the TI and MI layers, which exerts a spin torque on the magnetization of the MI layer to reverse it. The read unit has an MTJ exchange coupled to the MI layer that allows reading the information in the MI layer. The polarity of the output voltage can be changed on-the-fly by changing the polarity of the voltages $V^+$ and $V^-$ on the MTJ stack, allowing both inverting and non-inverting logic to be realized using the same primitive/layout. Typical material systems are, TI: Bi$_2$Se$_3$/(Bi$_\mathrm{x}$Sb$_\mathrm{1-x}$)$_2$Te$_3$, MI: Y$_3$Fe$_5$O$_{12}$, (Ni$_{0.65}$Zn$_{0.35}$)(Al$_{0.2}$Fe$_{0.8}$)O$_4$, BaFe$_{12}$O$_{19}$, Tm$_3$Fe$_5$O$_{12}$, MTJ: CoFeB-MgO-CoFeB/Ru/CoFe/IrMn (SAF), wires: metallic or semiconducting nano-interconnects with effective resistivity $<$ 100 $\mu\Omega$cm.}
\label{fig:device}
\vspace{-10pt}
\end{figure}

As the integrated Berry curvature of the full band does not vanish for a TI, and the spin Hall conductivity is directly related to the total Berry curvature of the filled states of the TI, even without any carriers near the chemical potential in bulk, the spin Hall conductivity does not vanish.  This characteristic clearly identifies the spin current involved as non-dissipative until it encounters other regions, such as an interface. Here we take advantage of this localized effect to drive vTOPSS shown in Fig.~\ref{fig:device}.
The device relies on the accumulation of spins at  an interface, originating from the voltage ($V_\mathrm{in}$) applied to the TI.  The spin Hall conductivity for a TI can be as large as (or larger than) that of a large SOC metal, but the dissipative longitudinal charge current will vanish for the TI.  Thus, a TI provides the advantages of a large spin Hall conductivity, but without the intrinsic dissipation of a metallic material. The resulting spin current produced by the electric field on the TI generates a torque on the spin in the magnetic material through exchange coupling or anti-damping torque. In the case of effective exchange coupling, the torque forces the magnetization to precess and eventually reverse.

The spin-current density created by applying an electric field $E_\mathrm{TI}$ is:
\begin{equation}
J_s=\sigma_{yx}E_\mathrm{TI}.
\end{equation}
The resulting magnetization dynamics of the ferromagnetic insulator can be described using the Landau-Lifshitz-Gilbert-Sloncewski equation in a macrospin limit~\cite{mayergoyz2009nonlinear}:
\begin{eqnarray}
\frac{1}{\gamma'}\frac{d \mathbf{m}}{d t} &=&-\mu_0 \mathbf{m}\times\mathbf{H}_\mathrm{eff}-\alpha \mu_0 \mathbf{m}\times\left(\mathbf{m}\times\mathbf{H}_\mathrm{eff}\right)\nonumber\\
&-& \underbrace{c_\mathrm{ex} j_s\mathbf{m}\times\mathbf{\hat{p}}}_{\text{Field-like torque}}+ \underbrace{j_s\mathbf{m}\times\left(\mathbf{m}\times\mathbf{\hat{p}}\right)}_{\text{Slonczewski torque}},
\label{Eqn:LLGS}
\end{eqnarray}
where $\mathbf{m}$ is a unit vector in the magnetization direction. $\gamma'=\gamma/(1+\alpha^2)$, $\gamma$ is the gyromagnetic ratio, $\mu_0$ is the vacuum permeability, and $\alpha$ is the Gilbert damping coefficient. The last two terms describe a spin torque from a spin-current polarized in a direction $\mathbf{\hat{p}}$, generally perpendicular to the electric field and in the plane of the TI/MI interface. $J_s=2M_s t_\mathrm{MI} j_s$, where $M_s$ is the magnetization of the MI layer, and $t_\mathrm{MI}$ is its thickness. (We assume a thin ferromagnetic insulator with area in contact with the topological insulator $A_\mathrm{int}$ and thickness $t_\mathrm{MI}$.)
The first spin-torque term describes the precession of the magnetization about the spin-polarization direction, with an exchange coupling parameter $c_\mathrm{ex}$. (For an estimate of this parameter see Ref.~\cite{Flatte2017}.)
This term is often referred to as a field-like interaction. The second spin-torque term characterizes the Slonczewski ``anti-damping'' torque, a torque that can oppose the dissipative term (the second term on the right hand side of the equation), leading to precessional magnetization dynamics and switching.

The effective field $\mathbf{H}_\mathrm{eff}$ characterizes the magnetic anisotropy of the free layer. For a uniaxial magnet with easy magnetization direction in the y-direction 
\begin{equation}
\mathbf{H}_\mathrm{eff}=2E_b m_y/(\mu_0 M_sV_\mathrm{MI}),
\end{equation}
 where $E_b$ is the energy barrier to magnetization reversal and $V_\mathrm{MI}$ is the volume of the MI layer.
%  ($V_\mathrm{MI}=A_\mathrm{int} t_\mathrm{MI}$). 
The magnetization switching mechanism depends on the orientation of the magnetic easy axis relative to the direction of spin-polarization $\mathbf{\hat{p}}$. When the two are orthogonal, the switching can occur due to precession about the spin-polarization direction and be very fast ($<100$ ps) \cite{Kent2004,Pinna2016,Rowlands2017}. However, typically precise electric pulse timing is required to ensure switching. When the spin-polarization is collinear with the easy-axis direction the switching is slower but the pulse time is not a critical parameter; in general, the write error rate decreases monotonically with either increasing pulse amplitude or duration~\cite{liu2014dynamics}. The electric field polarity determines the sense of reversal, i.e. from $m_y=1$ to $-1$ and vice-versa. The threshold spin current density for anti-damping spin-current switching follows from Eqn.~(\ref{Eqn:LLGS}), $J_{s,\mathrm{th}}=4\alpha E_b/A_\mathrm{int}$. The antidamping switching mechanism will be considered in the analysis presented in Sec.~\ref{sec:performance}.

As shown in Fig.~\ref{fig:device}, the 
readout in vTOPSS is accomplished by exchange coupling a small section of the MI layer (storing information) to the free layer of a magnetic tunnel junction (MTJ), which could operate with sub-100 mV supply voltages ($V^+/V^-$) to generate sufficient output voltage ($V_\mathrm{out}$) with intrinsic gain and the ability to fan-out. 
This separates the robust information storage aspect from the transduction within a hybrid magnetoelectric device, allowing one 
%This then allows one 
to probe the magnetization without disturbing the state.
%that follows.

% In general noise (e.g. thermal fluctuations) renders the switching stochastic.
% Need to decide how much we will discuss this or analyze this in the article.

\vspace{-10pt}

\vspace{-10pt}
\section{Performance Modeling} \label{sec:performance}
\vspace{-10pt}
In most spin-based devices, the operating speed is limited by the time it takes to reverse the magnetization of the metallic ferromagnetic layer, which is typically on the order of a nanosecond. %Moreover, large values of 
%electric current on the order of $(10^{7}-10^8)$ A/cm$^{2}$ are required to deterministically switch stable nanomagnets for which retention time is also an important criterion.
%with an energy barrier exceeding 30$k_BT$ to achieve sufficient retention time. 
%Here, $k_B$ is Boltzmann constant and $T$ denotes the operating temperature of the device. 
Spin-based devices using spin-Hall effect in heavy metals, such as Pt, Pd, W, require large electric fields in the heavy metal to generate sufficient electric current 
to cause STT switching of nanomagnets. 
%Large electric field requirement mainly stems from the low charge conductivity of heavy metals and their limited charge-to-spin conversion efficiency. The flow of electric current in existing spin-based devices not only causes excessive energy dissipation but could also lead to electromigration and reliability issues. 
The spin-based device, vTOPSS, takes advantages of the unique properties of TI and MI material systems to achieve the following criteria for energy-efficient logic applications: (i) non-volatility of operational states, (ii) fully voltage-driven switching of the MI magnetization with voltages $<$ 100 mV, (iii) absence of dissipative electric currents during the write process, and (iv) ultra-fast switching of the MI magnetization due to its low Gilbert damping. %In the next subsections, we quantify the 

In this section, analytic models of latency and energy dissipation of vTOPSS are presented followed by a comparison of metrics against those of existing spin-based and charge-based devices. Analytic models are obtained for a uniaxial MI layer subject to anti-damping STT resulting from spin accumulation at the TI-MI interface when the TI is subject to an external electric field. Multi-domain effects in the MI layer are neglected to arrive at closed form solutions of performance metrics that can provide insight into the device limits and opportunities. 

\vspace{-10pt}
\subsection{Device Latency}
\vspace{-10pt}
To estimate vTOPSS latency, the rate of spin accumulation at the TI-MI interface must be calculated. For a given 
electric field ($E_\mathrm{TI}$) and spin Hall conductivity ($\sigma_\mathrm{SHC}$) of the TI layer, the accumulation rate of interface spins is given as
\begin{eqnarray}
\frac{dn_\mathrm{spins}}{dt} = \frac{\sigma_\mathrm{SHC}}{\hbar/2}E_\mathrm{TI} = \frac{\sigma_\mathrm{SHC}}{\hbar/2}\frac{V_\mathrm{in}}{W}, 
\label{eq:nspins_int_rate}
\end{eqnarray} 
where 
$V_\mathrm{in}$ is the voltage applied across the TI layer, and $W$ is the width of the TI layer, measured along y-axis in Fig.~\ref{fig:device}.
%In Fig.~\ref{fig:device}, $W$ is measured along $\hat{y}$-axis.
For a given efficiency, $\varepsilon$, of coupling of spins at the TI-MI interface and the magnetic moment of the MI layer, the following condition is satisfied:
\begin{equation}
N_\mathrm{spins,MI} = \varepsilon n_\mathrm{spins}\mathcal{A}_\mathrm{int}.
\label{eq:Nspins_MI1}
\end{equation}
Here, $N_\mathrm{spins,MI}$ is the total number of spins in the MI layer subject to spin torque due to the interface spin accumulation, and $\mathcal{A}_\mathrm{int}$ is the interface cross-sectional area.
The total number of spins in a magnetic body is given as
\begin{eqnarray}
N_\mathrm{spins,MI} = \frac{M_s V_\mathrm{MI}}{\mu_B} = \frac{2E_b}{\mu_B H_K},
\label{eq:Nspins_MI}
\end{eqnarray} 
where $H_K$ is the anisotropy field of the MI layer, 
and $\mu_B = 9.3\times 10^{-24}$ J/T is Bohr magneton. Assuming anti-damping switching of the MI layer in the ballistic limit ($J_\mathrm{MI} \gg J_\mathrm{th}$), the rate of spin accumulation at the interface will balance the rate of magnetization reversal of the MI layer. Here, $J_\mathrm{MI}$ is the input spin current density in the MI layer, while $J_{\mathrm{th}}$ is the threshold spin current density required for STT-induced magnetization reversal. In this case, the reversal time, $\tau$, of the MI layer is~\cite{bedau2010spin} 
\begin{eqnarray}
\tau = \frac{N_\mathrm{spins,MI}}{\varepsilon \mathcal{A}_\mathrm{int}dn_\mathrm{spins}/dt}.
\label{eq:tau1}
\end{eqnarray}
Considering that the TI and MI widths are identical and using Eqn.~(\ref{eq:nspins_int_rate}) and Eqn.~(\ref{eq:Nspins_MI}), the above equation simplifies to
\begin{eqnarray}
\tau = \frac{2E_b}{\mu_B H_K \varepsilon L\left(\frac{\sigma_\mathrm{SHC}}{\hbar/2}\right) V_\mathrm{in}},
\label{eq:tau2}
\end{eqnarray}
where $L$ is the length of the TI layer (measured along x-axis in Fig.~\ref{fig:device}).
This equation shows that for fixed MI properties and switching voltage, the device latency is inversely proportional to the length scale. 
An increase in $L$ while fixing $E_b$ and $H_K$ values requires reducing the MI layer thickness. Therefore, the TI-MI interface area ($\mathcal{A}_\mathrm{int} = WL$) increases for the same volume of the MI layer, which increases the interface spin accumulation and the strength of STT acting on the MI layer.
%Since spin accumulation is proportional to the cross-sectional area,  
%increases the interface area, $\mathcal{A}_\mathrm{int}$ per 
%For fixed $E_b$ and $H_K$ values, increasing $L$ necessitates reducing the thickness to keep the number of spins in the MI layer fixed. 
%This scaling results from the fact that spin accumulation increases as the interface cross-sectional area increases, which increases the strength of STT acting on the MI layer. 

The device delay can also be reduced by lowering the MI energy barrier, $E_b$; however, this comes at the cost of reduced thermal stability of the MI layer.  
An increase in $\sigma_\mathrm{SHC}$ of the TI layer and the spin-coupling efficiency are particularly beneficial toward reducing the device latency. 
While the total latency of the device must include the time needed to charge/discharge the device capacitance (sum of interconnect and TI input capacitance), our analysis presented in Sec.~\ref{sec:energy} shows that the dominant time constant is due to the rate of spin accumulation at the TI-MI interface.
% that the total latency of the device  
 
Apart from the device geometry and material properties, a critical parameter affecting the device dynamics is the switching voltage, $V_\mathrm{in}$. This voltage must be enough to ensure that the spin current input to the MI layer exceeds the critical spin current ($J_\mathrm{th}$) for deterministic reversal. 
%In the case of uniaxial MI layer, the critical spin current is %given as
%$J_\mathrm{th} = 4 \alpha E_b/\mathcal{A}_\mathrm{int}$. 
%The Gilbert damping coefficient, $\alpha$, can be as low as 10$^{-4}$ for MI layers. This value is significantly lower than
%that achieved in metallic ferromagnets. 
Lower-$\alpha$ MI materials are advantageous to reduce $J_\mathrm{th}$ and permit low-power spin-based devices. 
%By cutting down the value of $J_\mathrm{th}$,
%lower-$\alpha$ magnetic materials are advantageous for practical implementation of low-power spin-based devices. 
The minimum switching voltage is found by considering $J_\mathrm{MI} = J_\mathrm{th} = d(N_\mathrm{spins,MI}/\mathcal{A}_\mathrm{int})/dt$. Using Eqn. (\ref{eq:nspins_int_rate}) and Eqn. (\ref{eq:Nspins_MI1}), we obtain
\begin{eqnarray}
V_\mathrm{in}^\mathrm{min} = \frac{2 \alpha E_b}{\varepsilon L \sigma_\mathrm{SHC}}.
\label{eq:Vmin}
\end{eqnarray}
Assuming $V_\mathrm{in} = \xi V_\mathrm{in}^\mathrm{min}$ ($\xi > 1$) and substituting in Eqn.~(\ref{eq:tau2}), we see that $\tau = \tau_D/(2\xi)$, where $\tau_D = 1/(\gamma H_K \alpha)$ is the natural time scale for the dynamics of uniaxial nanomagnets subject to anti-damping torque.
Considering $\alpha$ = 10$^{-4}$, $E_b$ = 30$kT$ ($kT$ = 25.8 meV at room temperature), $\varepsilon$ = 0.1, $L$ = 10 nm, $\sigma_{SHC}$ = 1000$\hbar/2e$ $\mathrm{\Omega}^{-1}$cm$^{-1}$, $V_\mathrm{in}^\mathrm{min}$ $\approx$ 0.75 mV. For $V_\mathrm{in}$ = 100 mV, corresponding to $\xi$ = 133, and $H_K$ = 0.1 T, $\tau$ $\approx$ 210 ps.

\vspace{-10pt}
\subsection{Minimum read voltage and energy dissipation} \label{sec:energy}
\vspace{-10pt}
In the vTOPSS device, the read unit is an MTJ stack coupled to the MI layer as depicted in Fig.~\ref{fig:device}. The read voltages are labeled as $V^+$ and $V^-$ and have the same magnitude but opposite polarity. 
That is, $V^+ = V_\mathrm{read}$ and $V^- = -V_\mathrm{read}$, where $V_\mathrm{read}$ is the magnitude of the voltage applied to the MTJ to read the magnetization state of the MI layer.
The supply voltages are clocked such that the writing and reading of a given logic stage happen in concurrent cycles.  
The voltage generated at the output node $V_\mathrm{out}$ of the $n^\mathrm{th}$ stage drives the write unit of $(n+1)^\mathrm{th}$ stage. The output voltage must meet the $V_\mathrm{in}^\mathrm{min}$ criterion in Eqn. (\ref{eq:Vmin}). 
\begin{figure}
\centering
\includegraphics[width=3.5in]{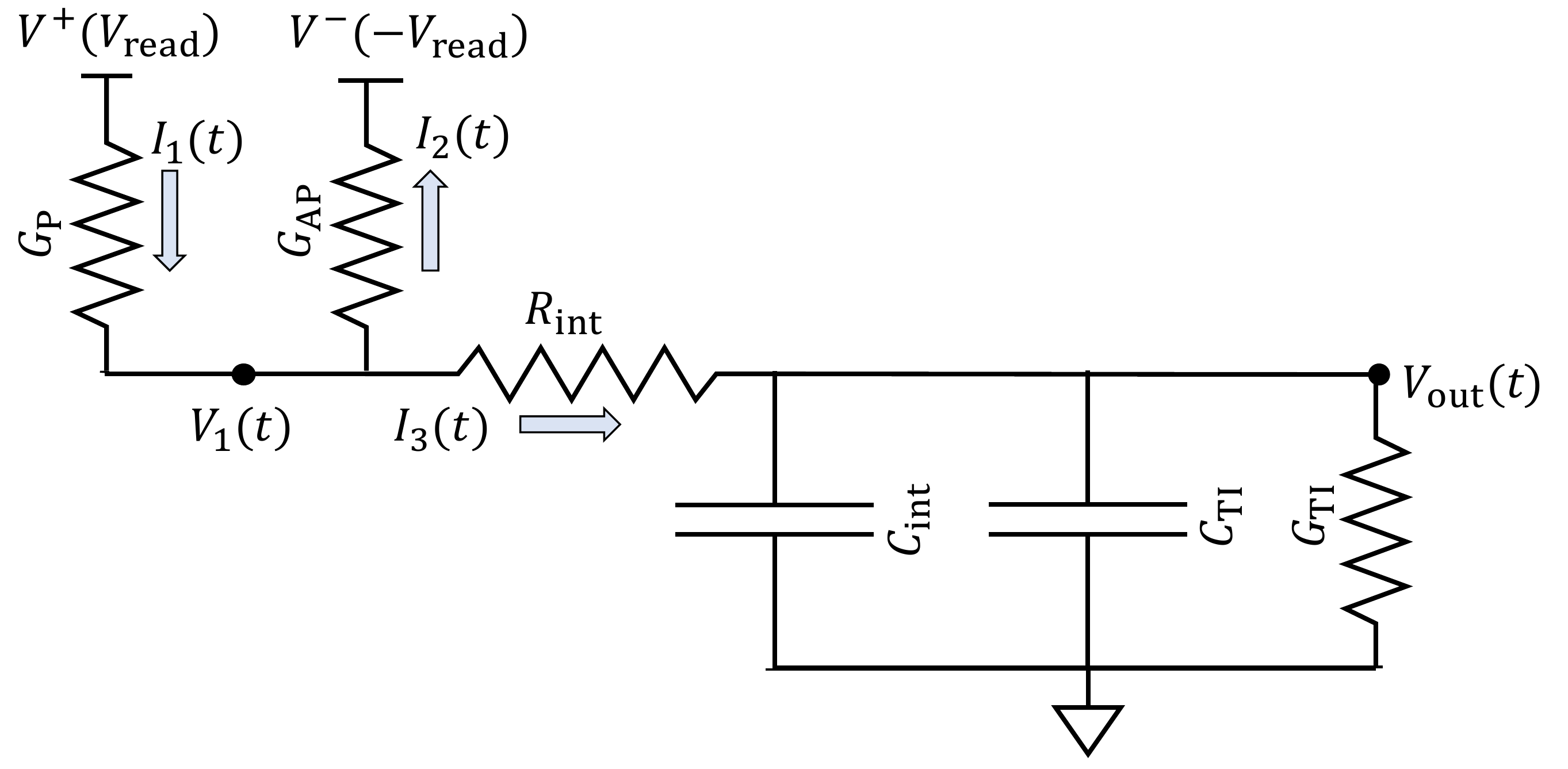} 
\caption{Equivalent electrical circuit of the read unit of a stage driving the write unit of the following stage. The MTJ stack conductances are given as $G_\mathrm{P}$ (parallel) and $G_\mathrm{AP}$ (anti-parallel). The total interconnect resistance and capacitance are $R_\mathrm{int}$ and $C_\mathrm{int}$. The TI layer is modeled as a leaky capacitor with capacitance $C_\mathrm{TI}$ and shunt conductance $G_\mathrm{TI}$. The MTJ read voltages have the same magnitude but opposite polarity.}
\label{fig:eq_ckt}
\vspace{-10pt}
\end{figure}

The equivalent circuit model of vTOPSS is shown in Fig.~\ref{fig:eq_ckt}.
The interconnect is modeled as a lumped RC network with $R_\mathrm{int}$ and $C_\mathrm{int}$ representing the total interconnect resistance and capacitance, respectively~\footnote{While a lumped interconnect model provides a pessimistic value of interconnect latency, we choose this model for its simplicity and ease of analytic calculations. Moreoever, for vTOPSS interconnect latency is significantly smaller than that associated with spin accumulation and reversal of the MI layer. Therefore, the error introduced due to a lumped model is negligible.}.
For a given interconnect length of $L_\mathrm{int}$, $R_\mathrm{int} = r_\mathrm{int}L_\mathrm{int}$ and $C_\mathrm{int} = c_\mathrm{int}L_\mathrm{int}$, where $r_\mathrm{int}$ and $c_\mathrm{int}$ are the per-unit-length interconnect resistance and capacitance, respectively. 
The conductances of the parallel and anti-parallel configuration of the free and fixed layer in the MTJ stack are denoted as $G_\mathrm{P}$ and $G_\mathrm{AP}$, respectively. These conductances are typically determined from the tunneling magnetoresistance (TMR) and resistance-area (RA) product measurements of the MTJ structure. TMR is given as $(G_\mathrm{P}-G_\mathrm{AP})/G_\mathrm{P}$, while RA is given as $\mathcal{A}_\mathrm{MTJ}/(G_\mathrm{P}+G_\mathrm{AP})$, where $\mathcal{A}_\mathrm{MTJ}$ is the cross-sectional area of the MTJ stack. For all results reported in this paper, $\mathcal{A}_\mathrm{MTJ}$ = $\mathcal{A}_\mathrm{int}$, unless otherwise specified.

The capacitance of the TI layer is $C_\mathrm{TI}$, while the leakage of electric current through the TI is modeled using the leakage conductance $G_\mathrm{TI}$.
The TI capacitance is given as $C_\mathrm{TI} = \epsilon_0\epsilon_r\mathcal{A}_\mathrm{int}/W$, where $\epsilon_0 = 8.85\times 10^{-12}$ F/m and $\epsilon_r$ is the static relative dielectric permittivity of the TI layer. For Bi$_{2}$Se$_{3}$, $\epsilon_r \approx 110$~\cite{clasen1998non}.
The leakage conductance $G_\mathrm{TI} = G_\mathrm{sheet} L/W$, where $G_\mathrm{sheet}$ = $en_\mathrm{s}\mu$~\cite{brahlek2015transport}. Here, $n_\mathrm{s}$ and $\mu$ correspond to the density and the effective mobility of surface carriers, respectively. 
 
The leakage in the TI layer results from the conductance of topologically trivial and non-trivial surface states as well as the bulk conductivity resulting from unavoidable self-doping effects~\cite{de2017coexistence}. 
Attempts to suppress bulk conductivity include thinning the TI layer until the surface contribution dominates or utilizing compensation doping to suppress free carriers in the bulk. 
For example, in Ref.~\cite{brahlek2014emergence}, copper doping is used in Bi$_{2}$Se$_{3}$ films to fully suppress bulk states and decouple the surface states in samples as thin as 20 nm. A sheet resistance of $\approx$ 1000 $\Omega/\Box$ at room temperature (300 K) is experimentally measured in a 20-nm thick Bi$_{2}$Se$_{3}$ film, while the sheet resistance increases to 1400 $\Omega/\Box$ and 3000 $\Omega/\Box$ in film thicknesses of 10 nm and 2 nm, respectively, in the same sample.
More recently, sheet resistances on the order of 10's of k$\Omega/\Box$ have been experimentally achieved at room temperature in 5-60 nm thick Bi$_{2}$Se$_{3}$ films grown on insulating In$_2$Se$_3$/(Bi$_{0.5}$In$_{0.5}$)$_2$Se$_3$ buffer layer on sapphire substrates~\cite{salehi2016finite}.

To obtain the energy dissipation of vTOPSS, Kirchoff's laws are first solved in the circuit shown in Fig.~\ref{fig:eq_ckt}, which gives the following time-domain response of output voltage:
%Solving KCL and KVL in this circuit gives the following time-domain response of charging the total output capacitance ($C_{out} = C_{int}+C_{TI}$):
\begin{subequations}
\begin{equation}
%\begin{split}
%V_{out}(t) = V_{f}\left(1-\exp\left(-\frac{t}{\tau_{eq}}\right)\right) +V_{i}\exp\left(-\frac{t}{\tau_{eq}}\right),
V_\mathrm{out}(t) = V_\mathrm{f}\left(1-e^{-\frac{t}{\tau_\mathrm{eq}}}\right) +V_\mathrm{i}e^{-\frac{t}{\tau_\mathrm{eq}}},
%\end{split}
%\label{eq:Vout_time}
\end{equation}
%\begin{widetext}
%\[
%V_{out,steady} = \frac{V^+G_P+V^-G_{AP}}{G_P+G_{AP}+R_{int}G_{TI}\left(G_{P}+G_{AP}\right)} = \frac{(G_P-G_{AP})V_{read}}{G_P+G_{AP}+R_{int}G_{TI}\left(G_{P}+G_{AP}\right)},\]
%\end{widetext}
%\begin{equation}
%\parbox[t]{\displaywidth}{
%    \raggedright
%$V_{out,steady} = \frac{V^+G_P+V^-G_{AP}}{G_P+G_{AP}+R_{int}G_{TI}\left(G_{P}+G_{AP}\right)} = \frac{(G_P-G_{AP})V_{read}}{G_P+G_{AP}+R_{int}G_{TI}\left(G_{P}+G_{AP}\right)}$}
%\end{equation}
\begin{equation}
V_\mathrm{f} = \frac{(G_\mathrm{P}-G_\mathrm{AP})V_\mathrm{read}}{G_\mathrm{P}+G_\mathrm{AP}+R_\mathrm{int}G_\mathrm{TI}\left(G_\mathrm{P}+G_\mathrm{AP}\right)},
\end{equation}
\begin{equation}
\tau_\mathrm{eq} = \frac{\left(1+R_\mathrm{int}(G_\mathrm{P}+G_\mathrm{AP})\right)C_\mathrm{out}}{G_\mathrm{P}+G_\mathrm{AP}+R_\mathrm{int}G_\mathrm{TI}\left(G_\mathrm{P}+G_\mathrm{AP}\right)}.
\end{equation}
\label{eq:Vout}
\end{subequations}
\hspace{-7pt}
Here, $V_\mathrm{f}$ and $V_\mathrm{i}$ are the final and initial voltages, respectively, at the output node. 
At the end of the read/write cycle, the voltage $V_\mathrm{out}$ is reset to 0 V.
%$V_{init}$ is the initial voltage at the output node. 
%After every read/write cycle, the voltage at $V_{out}$ i. 
Therefore, for all results presented in this paper, $V_\mathrm{i}$ = 0 V.
The minimum read voltage 
required on the MTJ stack to ensure correct functionality is obtained by equating Eqn. (\ref{eq:Vmin}) and Eqn. (\ref{eq:Vout}). Assuming that the read pulse duration is significantly greater than $\tau_\mathrm{eq}$, $V_\mathrm{read}^\mathrm{min}$ is given as 
\begin{equation}
V_\mathrm{read}^\mathrm{min} = \frac{2\alpha E_b}{\varepsilon L \sigma_\mathrm{SHC}}\left[\frac{(G_\mathrm{P}+G_\mathrm{AP})(1+R_\mathrm{int}G_\mathrm{TI})}{G_\mathrm{P}-G_\mathrm{AP}}\right].
\label{eq:Vread_min}
\end{equation}

\begin{figure}
%\vspace{-4ex}
\centering
\includegraphics[width=3.25in]{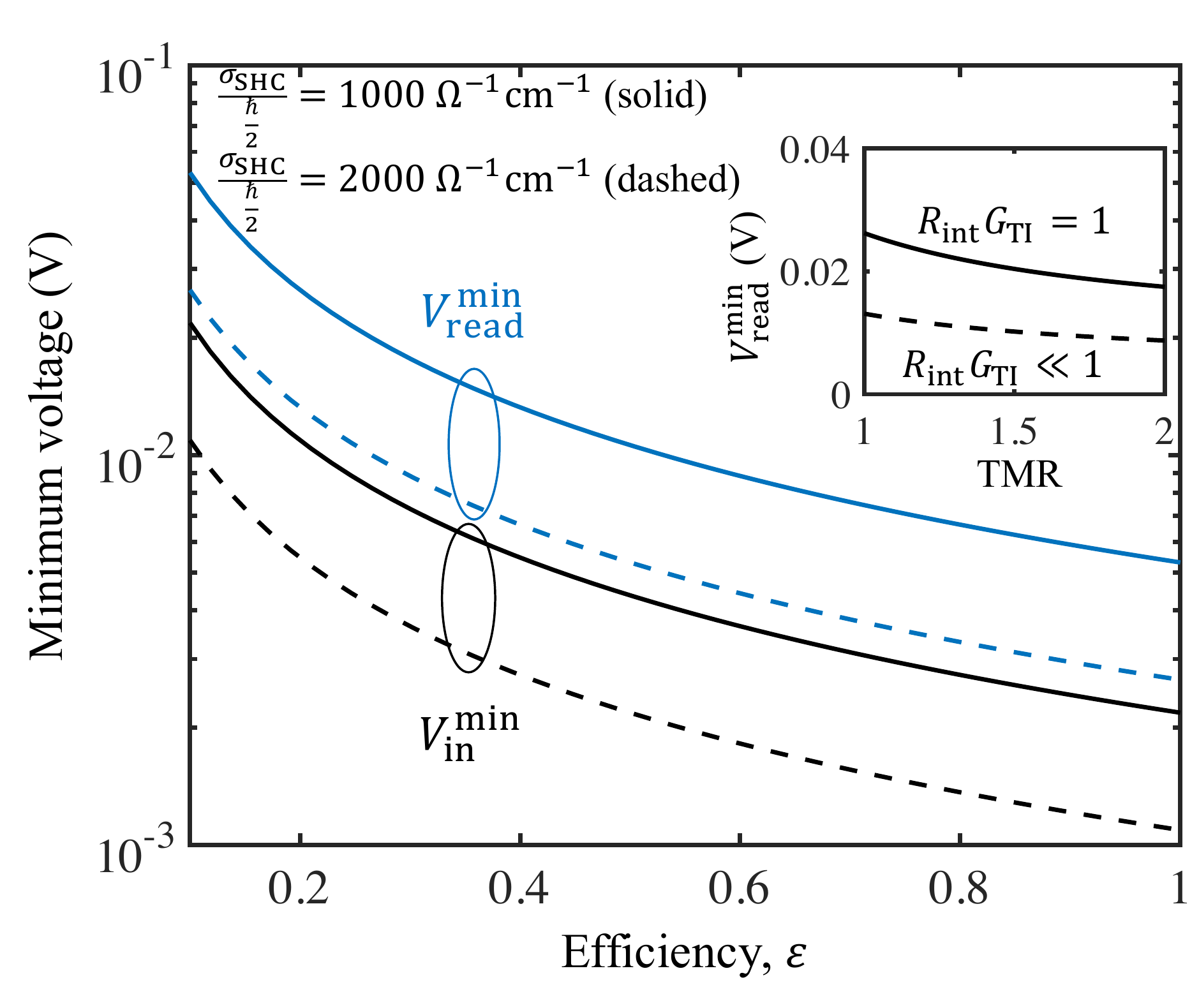} 
\vspace{-4ex}
\caption{Minimum input and read voltages versus the spin coupling efficiency in vTOPSS for various values of the spin Hall conductivity of the TI layer ($\sigma_\mathrm{SHC}$). Simulation parameters: $\alpha$ = 3$\times 10^{-3}$, $E_b$ = 30$kT$, TMR = 1.4, $L$ = 10 nm, $W$ = 100 nm, $R_\mathrm{int}G_\mathrm{TI}\ll 1$. Inset plot shows the scaling of minimum read voltage with TMR for various $R_\mathrm{int}G_\mathrm{TI}$ product values. Note that the area of the MTJ and the RA product values do not affect the value of $V_\mathrm{read}^\mathrm{min}$ and $V_\mathrm{in}^\mathrm{min}$.}
\label{fig:min_volt}
\vspace{-10pt}
\end{figure}

In Fig.~\ref{fig:min_volt}, the minimum read voltage ($V_\mathrm{read}^\mathrm{min}$) and the minimum input voltage ($V_\mathrm{in}^\mathrm{min}$) required for magnetization reversal are plotted as functions of the efficiency of spin coupling at TI-MI interface for various values of the spin Hall conductivity. The effect of TMR on $V_\mathrm{read}^\mathrm{min}$ is examined in the inset plot. 
Our results show that sub-20 mV input voltages corresponding to sub-50 mV read voltages will enable device functionality even when the coupling efficiency is as low as 10\%. 
As expected, an improvement in coupling efficiency, spin Hall conductivity, and the TMR can lower the required supply voltage to only a few milli-volts. Such a low switching voltage to reverse the magnetization of magnetic materials is key toward enabling  ultra-low-energy operation using vTOPSS.

The total energy dissipation consists of the energy required to charge and discharge the output node voltage, $V_\mathrm{out}$, and the direct path conduction between $V^+$ and $V^-$. 
Assuming that the read phase lasts for a time duration of 
%If the voltages on the MTJ during the read phase are applied for a time duration of 
$\tau_\mathrm{pulse}$, the energy supplied by the voltage $V^+$ is given by the following integral:
%$E_{read} = \int_0^{\tau_{pulse}}{dt I_1(t) V^+} = \int_0^{\tau_{pulse}}{dt (I_2(t)+I_3(t))V^+ }$.
%The instantaneous power dissipation is the time derivative of the energy supplied by the read voltage and is given as 
%\begin{subequations}
\begin{equation}
E_\mathrm{read} = \int_0^{\tau_\mathrm{pulse}}{dt I_1(t) V^+} = \int_0^{\tau_\mathrm{pulse}}{dt (I_2(t)+I_3(t))V^+ },
\end{equation}
%\begin{equation}
%I_2(t) = G_{AP}(V_1(t)-V^-),
%\end{equation}
%\begin{equation}
%\begin{split}
%I_3(t) = \frac{(V_1(t)-V_{out}(t))}{R_{int}} = C_{out}\frac{dV_{out}(t)}{dt}+G_{TI}V_{out}(t).
%\end{split}
%\end{equation}
%\end{subequations}
where $I_j(t)$ ($j$ = 1,2,3) denotes the electrical current flowing in the $j^{\mathrm{th}}$ branches as shown in Fig.~\ref{fig:eq_ckt}, 
$I_2(t) = G_\mathrm{AP}(V_1(t)-V^-)$, and $I_3(t) = C_\mathrm{out}dV_\mathrm{out}(t)/dt+G_\mathrm{TI}V_\mathrm{out}(t)$.
Here $C_\mathrm{out} = C_\mathrm{TI}+C_\mathrm{int}$ is the net capacitive loading at the output node.
By substituting $V_1(t)$ in terms of $V_\mathrm{out}(t)$ and using Eqn. (\ref{eq:Vout}), the energy dissipation of the circuit is given per Eqn. (\ref{eq:Eread_total}).
The term $E_\mathrm{read,1}$ in Eqn. (\ref{eq:Eread_total}) is dominated by the energy dissipation due to MTJ leakage, while $E_\mathrm{read,2}$ is largely due to energy consumed in charging and discharging the output capacitance. The term $E_\mathrm{read,3}$ couples the energy dissipation due to the finite conductance of the TI layer and the MTJ. For $G_\mathrm{TI} \gg G_\mathrm{AP}$, the leakage through the TI layer dominates $E_\mathrm{read,3}$. 
%In this equation, the first term ($E_\mathrm{read,1}$) captures the effect of 

%
%found to be
%\begin{equation}
%\begin{split}
%E_{read} = V_{read}^2G_{AP}\tau_{pulse}+ \\ C_{out}V_{read}V_{out}(\tau_{pulse})\left[1+R_{int}G_{AP}\right] \\ + V_{read}V_{out,steady}\tau_{pulse}\left[G_{TI}+G_{AP}(1+R_{int}G_{TI})\right] \\ \left[1-\frac{\tau_{eq}}{\tau_{pulse}}+\frac{\tau_{eq}}{\tau_{pulse}}\exp\left(\frac{-\tau_{eq}}{\tau_{pulse}}\right)\right].
%\end{split}
%\end{equation}
The time to charge/discharge the output capacitance is $\approx \tau_\mathrm{eq}$.
For typical material parameters, $\tau_\mathrm{eq}$ is much smaller than the time 
required for spin accumulation and STT-driven reversal of the MI layer. 
For example, considering $r_\mathrm{int}$ = 1.25$\times 10^7$ $\Omega$/m, $L_\mathrm{int}$ = 100 nm, TMR = 1.4, RA = 4$\times 10^{-11}$ $\Omega$m$^2$, $\mathcal{A}_\mathrm{MTJ}$ = 333 nm$^2$, $C_\mathrm{out}$ = 0.1 fF, $\tau_\mathrm{eq} \approx$ 13 ps. 
On the other hand, reversal delay of the MI layer is several hundreds of picoseconds indicating that charging/discharging of load capacitances through interconnects is not the limiting factor in vTOPSS technology (utilizing anti-damping switching of the MI layer).
This is contrary to CMOS technology in which local interconnects as short as a few gate pitches dominate the circuit delay~\cite{rakheja2010interconnects}.  
%Therefore, $\tau_\mathrm{pulse} >> \tau_\mathrm{eq}$.

For $\tau_\mathrm{pulse} \gg \tau_\mathrm{eq}$, a lower bound on the energy dissipation may be obtained by setting $\tau_\mathrm{pulse} = \tau$ (defined in Eqn.~(\ref{eq:tau2})), $V_\mathrm{read} = V_\mathrm{read}^\mathrm{min}$ (defined in Eqn.~(\ref{eq:Vread_min})), and $V_\mathrm{f} = V_\mathrm{in}^\mathrm{min}$ (defined in Eqn.~(\ref{eq:Vmin})). 
%This yields 
%\begin{equation}
%\begin{split}
%E_{read}^{min} = k(V_{in}^{min})^2\left[\tau\left(G_{TI}+G_{AP}(1+k+R_{int}G_{TI})\right) + \\ C_{out}(1+R_{int}G_{AP})\right],
%\end{split}
%\end{equation}
%%$E_{read}^{min}$ = $k(V_{in}^{min})^2\left[\tau\left(G_{TI}+G_{AP}(1+k+R_{int}G_{TI})\right) + C_{out}(1+R_{int}G_{AP})\right]$, 
%where $k = \left[(G_P+G_{AP})(1+R_{int}G_{TI})\right]/(G_P-G_{AP})$.
%The steady state voltage $V_{out,steady} = V_{DD} \Delta G/G$, where $\Delta G = (G_P-G_{AP})$ and $G = (G_P + G_{AP})$. $G_P$ and $G_{AP}$ are the conductances of the parallel and anti-parallel configuration, respectively, of the free and fixed layer in the MTJ stack.
%For a given pulse width of the voltage ($\tau_{pulse}$)

\begin{widetext}
\begin{equation}
\begin{split}
E_\mathrm{read} = \underbrace{V_\mathrm{read}^2G_{AP}\tau_\mathrm{pulse}}_{E_\mathrm{read,1}}+ \underbrace{C_\mathrm{out}V_\mathrm{read}V_\mathrm{out}(\tau_\mathrm{pulse})\left[1+R_\mathrm{int}G_\mathrm{AP}\right]}_{E_\mathrm{read,2}} + \\  \underbrace{V_\mathrm{read}V_\mathrm{f}\tau_\mathrm{pulse}\left[G_\mathrm{TI}+G_\mathrm{AP}(1+R_\mathrm{int}G_\mathrm{TI})\right] \left[1-\frac{\tau_\mathrm{eq}}{\tau_\mathrm{pulse}}+\frac{\tau_\mathrm{eq}}{\tau_\mathrm{pulse}}e^{\frac{-\tau_\mathrm{eq}}{\tau_\mathrm{pulse}}}\right]}_{E_\mathrm{read,3}}.
\end{split}
\label{eq:Eread_total}
\end{equation}
\end{widetext}

\vspace{-10pt}
\section{Performance Benchmarking}
\vspace{-10pt}
To benchmark the performance of vTOPSS against CMOS and existing spin-based devices, we first study the impact of device design on vTOPSS latency, energy, and energy-delay product (EDP). 
For CMOS logic at the 2020 ITRS technology node, the effect of local interconnects (copper/low-$\kappa$) on the performance metrics is considered.
%metrics are obtained for both with and without interconnect considerations to quantify the impact of interconnects on the overall performance of CMOS logic. 
We also identify a set of optimal material parameters of vTOPSS to achieve an EDP on the order of $10^{-29}$ Js for it to be competitive against CMOS logic. 
%is identified to lower the EDP of vTOPSS below $10^{-29}$ J.s to be competitive against low-power CMOS design. 

\vspace{-10pt}
\subsection{Energy and latency}
\begin{figure}[h!]
\centering
\includegraphics[width=3.25in]{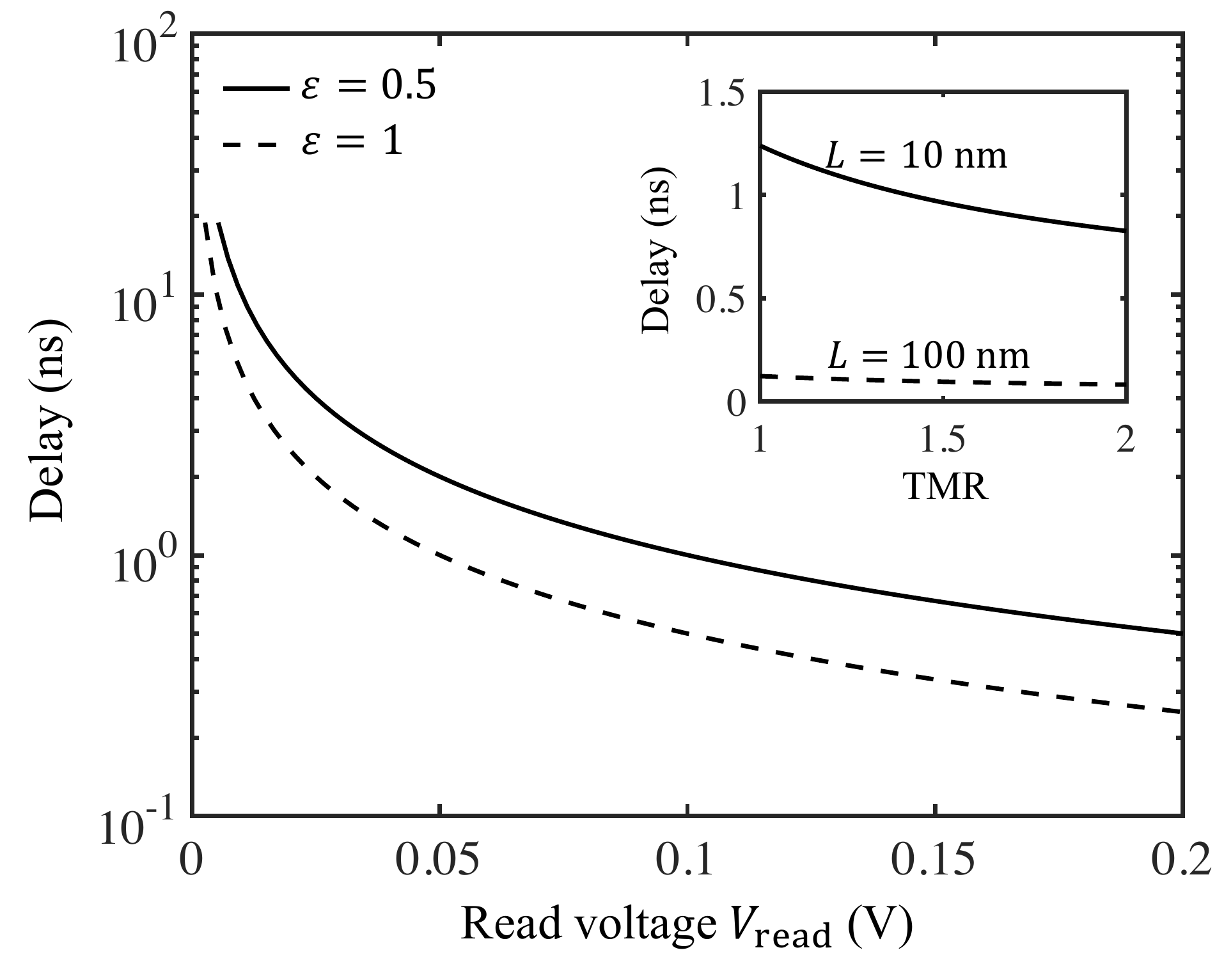} 
\caption{Delay of vTOPSS versus read voltage. Inset shows the impact of TMR on the delay for various values of the TI length. Simulation parameters are: $\sigma_\mathrm{SHC}$ = 2000$\hbar/2e$ $\mathrm{\Omega}^{-1}$cm$^{-1}$, $R_\mathrm{sheet}$ = 10 k$\Omega/\Box$, $\alpha$ = 3$\times 10^{-3}$, $E_b$ = 30$kT$, $H_K$ = 0.1 T, RA product = 100 $\Omega\mu$m$^2$, TMR = 1.4, $\mathcal{A}_\mathrm{MTJ}$ = $\mathcal{A}_\mathrm{int}$, $c_\mathrm{int}$ = 1.6 pF/cm, $r_\mathrm{int}$ = 1.25$\times 10^7$ $\mathrm{\Omega}$/m, $L_\mathrm{int}$ = 100 nm, $L$ = 10 nm, $W$ = 100 nm. For the inset plot, $V_\mathrm{read}$ = 100 mV, $\varepsilon$ = 1, while the TI-MI interface length ($L$) is varied.}
\label{fig:delay}
\vspace{-10pt}
\end{figure}

\vspace{-10pt}
Figure~\ref{fig:delay} shows the latency of vTOPSS plotted as a function of the read voltage. 
Simulation parameters are noted in the figure caption.
The dominant time constant in vTOPSS is the magnetization reversal delay, which can be reduced by increasing the read voltage on the MTJ or by improving the efficiency of coupling of TI-MI spins. For a TI length of 10 nm, read voltage of 50 mV, and $\varepsilon$ = 0.5, the latency of vTOPSS is $\approx$ 2 ns, while the latency reduces to $\approx$ 1 ns for perfect spin coupling efficiency. 
To ensure that the MI reversal delay is sub-1 ns, the area of the interface between the TI and MI layers must be increased. 
As shown in the inset plot of Fig.~\ref{fig:delay}, by increasing the length of the TI layer to 100 nm, a delay of only a few 100 ps for vTOPSS is achievable.
With all other parameters fixed, an increase in TMR increases the input voltage across the TI layer, reducing the latency of MI reversal. Another important material property is the spin Hall conductivity ($\sigma_\mathrm{SHC}$) of the TI layer--an increase in $\sigma_\mathrm{SHC}$ increases the interface spins available to drive the MI magnetization, thereby reducing its delay.
%Alternately, for a fixed interface area, an increase in spin Hall conductivity of the TI layer will also reduce the MI reversal delay.

\begin{figure}[h!]
\includegraphics[width=3.25in]{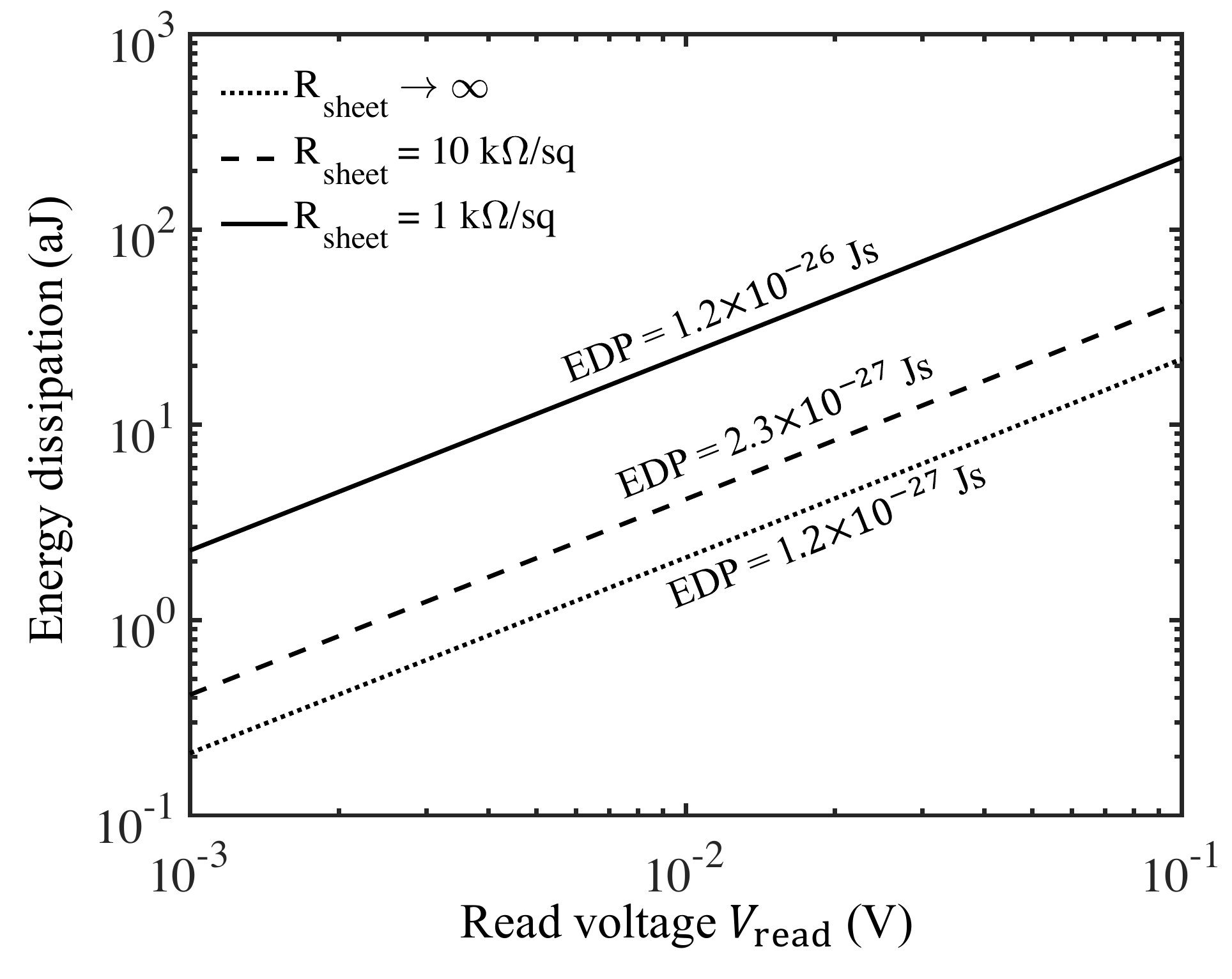}
\caption{Energy dissipation of vTOPSS as a function of the read voltage. TI-MI interface are $\mathcal{A}_\mathrm{int}$ = (100$\times$100) nm$^2$, and spin coupling efficiency, $\varepsilon$ = 1. Other simulation parameters are the same as those reported in Fig.~\ref{fig:delay}.}
\label{fig:energy}
\vspace{-10pt}
\end{figure}

\begin{figure}[h!]
\centering
\includegraphics[width=3.25in]{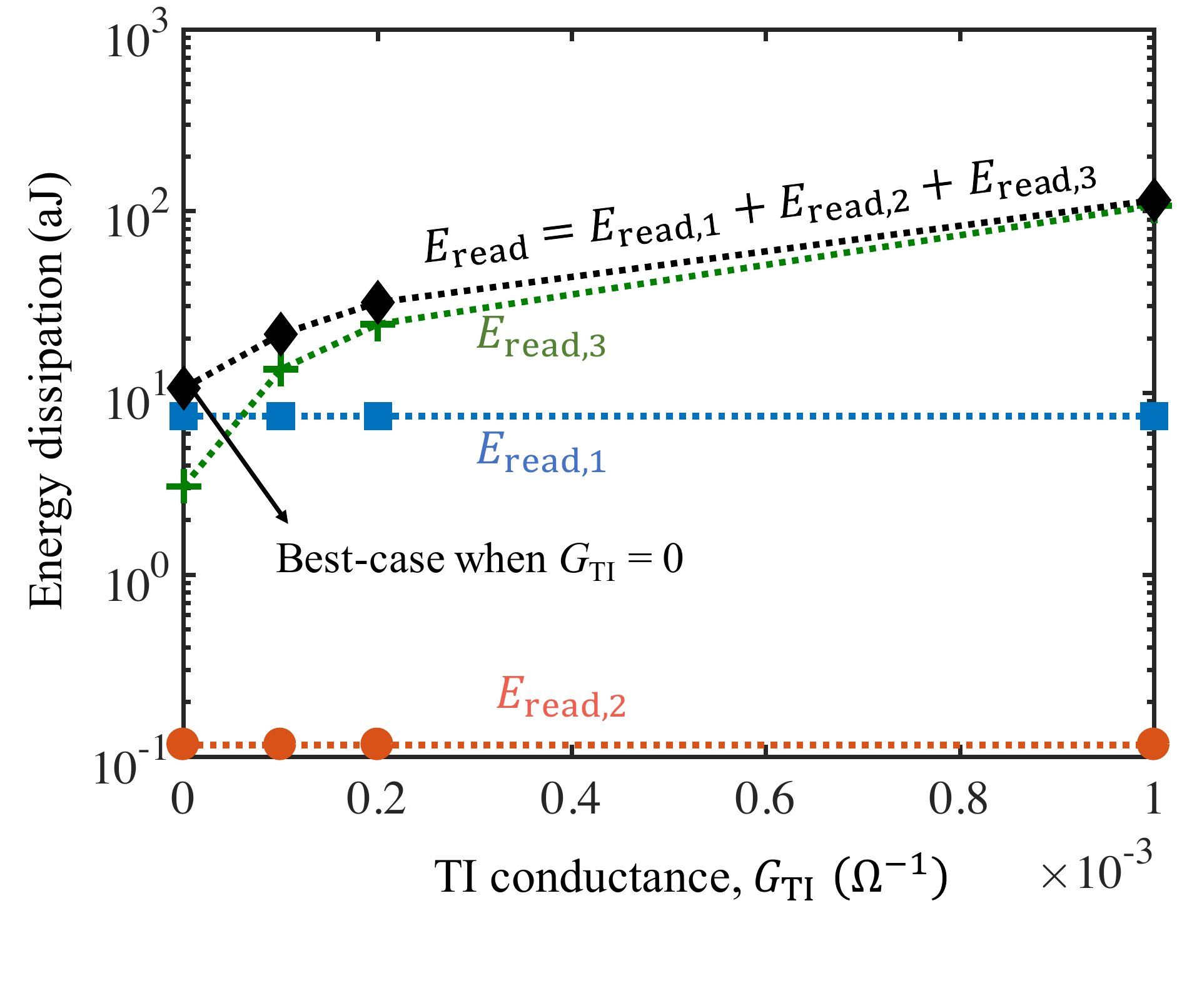} 
\vspace{-15pt}
\caption{Contribution of various terms to the energy dissipation as a function of the TI conductance. Simulation parameters are the same as in Fig.~\ref{fig:energy}.}
\label{fig:energy2}
\vspace{-10pt}
\end{figure}

Figure~\ref{fig:energy} shows the scaling of vTOPSS energy dissipation with the read voltage applied on the MTJ.
As the read voltage reduces the energy dissipation also reduces. However, the energy dissipation of vTOPSS decreases linearly with $V_\mathrm{read}$.  
At a read voltage of 50 mV, the energy dissipation of vTOPSS for infinite sheet resistance (gapped surface states and negligible bulk conductivity) is as low as 10 aJ. This energy dissipation increases by 12$\times$ for Bi$_2$Se$_3$ thin films with a sheet resistance of 1 k$\Omega/\Box$ measured at room temperature in Ref.~\cite{brahlek2014emergence}.
Figure~\ref{fig:energy2} shows the contribution of various terms in Eqn.~(\ref{eq:Eread_total}) to the overall vTOPSS energy dissipation. In the best-case scenario with negligible TI conductance, the energy dissipation is mainly dominated by the first term in Eqn. (\ref{eq:Eread_total}). This term signifies the importance of leakage through the MTJ stack and can be reduced by utilizing material systems with a much larger TMR ratio. As the leakage through the TI increases, energy dissipation begins to be dominated by the third term in Eqn.~(\ref{eq:Eread_total}).  
Note that even if $G_\mathrm{TI} \rightarrow$ 0, the contribution of the third term in Eqn.~(\ref{eq:Eread_total}) remains finite due to the presence of $G_\mathrm{AP}$. 
Finally, the contribution of the second term in Eqn.~(\ref{eq:Eread_total}) remains insignificant as long as $R_\mathrm{int}G_\mathrm{TI} \ll 1$. This condition is typically satisfied for local metallic interconnects considered for results in Fig.~\ref{fig:energy}.
The effect of interconnects on energy dissipation is further discussed in Sec.~\ref{sec:interconnect}.

\vspace{-10pt}
\subsection{Energy-delay product (EDP)}
\vspace{-10pt}
As shown in Figs.~\ref{fig:delay} and \ref{fig:energy}, there exists an energy-delay tradeoff in vTOPSS with respect to $V_\mathrm{read}$. However, this trade-off in vTOPSS has a subtle difference when compared against CMOS technology.
In CMOS logic, energy dissipation scales quadratically with the supply voltage ($E_\mathrm{CMOS} \propto V_{DD}^2$). 
In vTOPSS, however, energy dissipation displays a linear dependence on the read voltage. 
This is because leakage through the TI conductance and the MTJ conductance dominate vTOPSS energy (see results and discussion related to Fig.~\ref{fig:energy}). Since, $\tau_\mathrm{pulse} \approx \tau \propto V_\mathrm{read}^{-1}$ and $E_\mathrm{read} \propto V_\mathrm{read}$, a constant EDP with respect to $V_\mathrm{read}$ is obtained in vTOPSS, 
%This scaling generates a constant EDP in vTOPSS technology 
if all other design parameters were kept the same. For the results reported in Fig.~\ref{fig:energy}, the EDP increases from 1.2$\times 10^{-27}$ Js to 1.2$\times 10^{-26}$ Js as the sheet resistance of the TI layer reduces (or the TI conductance increases).

The EDP can be reduced by designing junctions that exhibit a large TMR--an increase in TMR at a fixed RA product value reduces both the switching delay and the energy dissipation.
The scaling of EDP with TMR in vTOPSS can be expressed as $\mathrm{EDP} \propto 1/\mathrm{TMR}^b$, where the exponent $b \geq 1$.
At the same time, there exists an optimal value of RA product that minimizes the energy dissipation and the EDP of vTOPSS. 
As long as $\tau_{eq}$ is negligible compared to the MI reversal delay, an increase in RA product reduces the energy dissipation, resulting in a concomitant reduction in the EDP of the device. Note that the MI reversal delay is not affected by the RA product. 
For STT-MRAM applications, a large RA product is undesirable as it increases the voltage required to switch the magnetization state via current-induced spin torques~\cite{chatterjee2011scalable}. In the case of vTOPSS technology, the MTJ cell is only used to generate a rather low output voltage that must be sufficient to switch the subsequent logic stage. As such, a large RA of the read unit on the MTJ stack may be beneficial
to the design of vTOPSS. 

\begin{figure}[h!]
\centering
\includegraphics[width=3.25in]{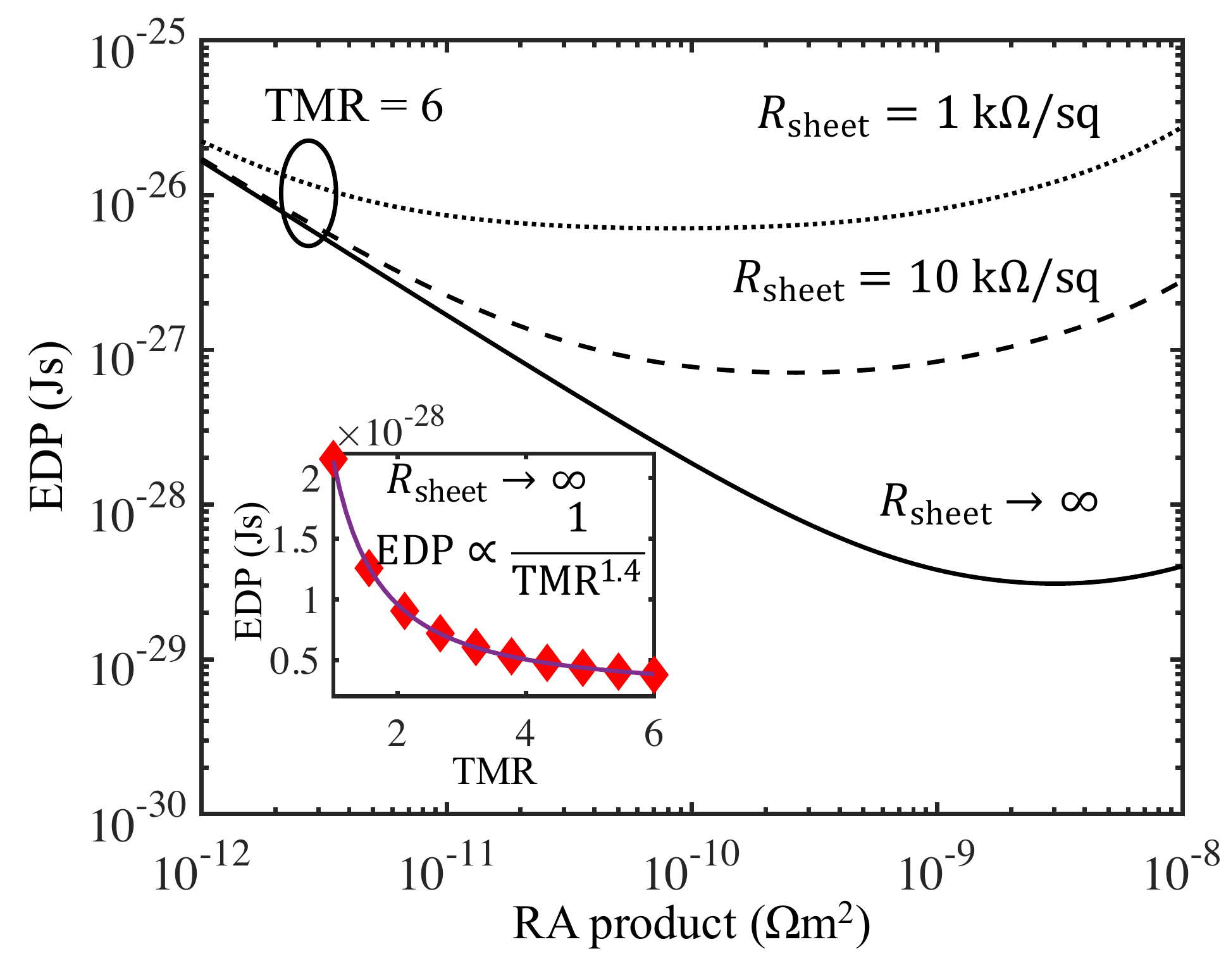} 
\caption{EDP of vTOPSS versus the RA product of the MTJ for various values of $R_\mathrm{sheet}$ of the TI layer at $\varepsilon$ = 1, TMR = 600\%, $V_\mathrm{read}$ = 50 mV. An optimal RA product value exists that minimizes the EDP of the device. The inset shows the scaling of EDP with TMR for all other material parameters fixed. Inset shows the scaling of EDP with TMR for RA = 10$^{-9}$ $\Omega$m$^2$. Other simulation parameters are same as those reported in Fig.~\ref{fig:delay}.}
\label{fig:EDP_3D}
\vspace{-10pt}
\end{figure}

As shown in Fig.~\ref{fig:EDP_3D}, EDP initially decreases with an increase in RA product. With further increase in RA, EDP exhibits the reverse trend and begins to increase. 
For RA $<$ RA$_\mathrm{opt}$ (optimal), the scaling of EDP with RA is expressed as $\mathrm{EDP} \propto \mathrm{RA}^{-1}$. 
Beyond the optimal value RA$_\mathrm{opt}$, unfortunately the delay associated with charging/discharging capacitive nodes becomes much larger than the MI reversal delay. 
The optimal value of RA depends on the material and geometry of the device. For the results shown in Fig.~\ref{fig:EDP_3D}, RA$_\mathrm{opt}$ decreases with a reduction in $R_\mathrm{sheet}$. Moreover, the EDP-RA contour becomes flatter around the optimal point as $R_\mathrm{sheet}$ reduces.
%an increase in RA product values up to 10$\Omega\mu$m$^2$ reduces the EDP significantly. 
Results show that at a TMR of 600\% and without any leakage through the TI, 
%RA$_\mathrm{opt}$, and no leakage through the TI, 
%RA product of 10 $\Omega\mu$m$^2$, 
the optimal EDP of vTOPSS is around $2\times 10^{-29}$ Js, which is comparable to the EDP of CMOS technology at the 2020 ITRS node (see discussion in Sec.~\ref{sec:cmos}). For the same parameters, the energy dissipation and the delay of vTOPSS are 0.23 aJ and 55 ps, respectively.  
In typical MTJs, TMR increases with increasing RA product,
which can be harnessed to improve the EDP of vTOPSS. In~\cite{ikeda2005dependence}, it is shown that a TMR of 600\% can be obtained with an RA product of $10^4$ $\mathrm{\Omega}\mathrm{\mu m}^2$ in CoFeB/MgO/CoFeB type MTJs by annealing the structure above 500$^\circ$C.

%\subsection{Comparison against existing spin-based devices}
\vspace{-10pt}
\subsection{Interconnect considerations}\label{sec:interconnect}
\begin{figure}[h!]
\centering
\includegraphics[width=3.25in]{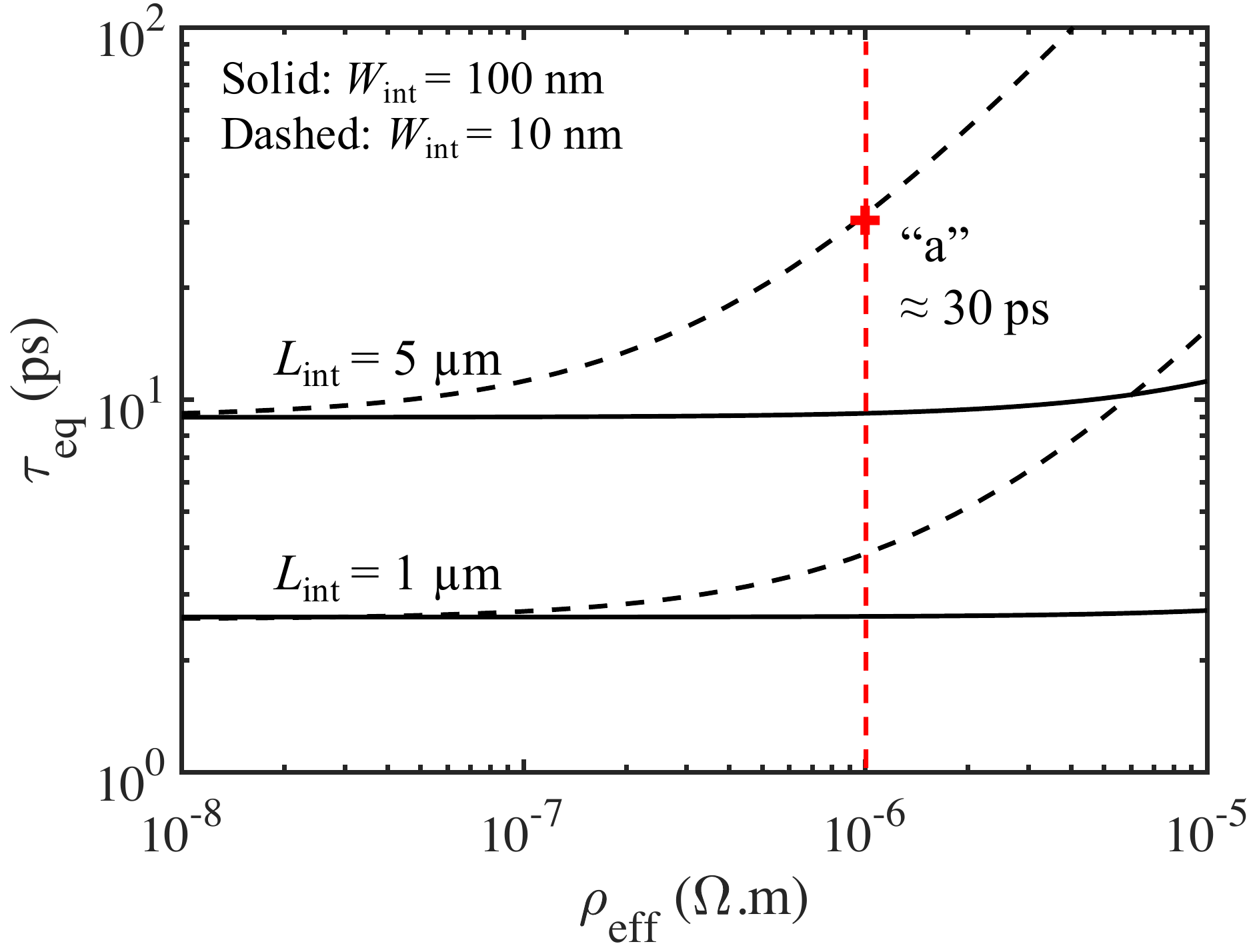} 
\caption{Effect of interconnect resistivity ($\rho_\mathrm{eff}$), length ($L_\mathrm{int}$), and width ($W_\mathrm{int}$) on the charging/discharging time constant of the output node voltage. For comparison, the MI reversal delay is on the order of few 100's of picoseconds. Simulation parameters are the same as those reported in the caption of Fig.~\ref{fig:delay}.}
\label{fig:int}
\end{figure}

\vspace{-10pt}
To transmit information between vTOPSS logic, conventional CMOS-compatible metallic interconnects can be used. Additionally, highly resistive nanowires with effective resistivity ($\rho_\mathrm{eff} \leq $ 100 $\mathrm{\mu\Omega}$cm) can also be used as the dominant component of delay is due to the MI reversal under the influence of anti-damping torque. As such, there exists a wide range of interconnect options, such as copper, ultra-scaled wires (wire width $\ll$ electron mean free path), doped semiconducting wires to design vTOPSS logic. The effect of interconnect resistivity on the the interconnect latency of vTOPSS for  various interconnect lengths is shown in Fig.~\ref{fig:int}. 
Even for interconnect resistivity of 100 $\mu\Omega$cm, the delay associated with charging/discharging the output node is $\approx$ 30 ps for interconnect length of 5 $\mu$m and width of 10 nm (point labeled as ``a'' in Fig.~\ref{fig:int}). For wider interconnects, the effect of resistivity increase on the charging/discharging time constant is negligible. For comparison, the reversal delay of the MI layer is few 100's of picoseconds and will dominate the overall latency of vTOPSS for interconnect length scales up to a few micrometers.  

\vspace{-10pt}
\subsection{Comparison against existing logic devices} \label{sec:cmos}
\vspace{-10pt}
The model used for computing the performance metrics of CMOS technology comprise a minimum-sized CMOS inverter driving a similarly sized load through copper/low-$\kappa$ interconnect.  
%with a length of 100 nm. CMOS device metrics are taken from the 2013 ITRS roadmap. 
Using the Elmore delay model, the delay of the CMOS circuit is given as~\cite{bakoglu1985optimal} 
\begin{equation}
\begin{split}
\tau_\mathrm{CMOS} = 0.69R_\mathrm{S}\left(C_\mathrm{S}+C_\mathrm{L}\right)+ \\ 0.69\left(R_\mathrm{S}c_\mathrm{int}+r_\mathrm{int}C_\mathrm{L}\right)L_\mathrm{int}+ 0.38r_\mathrm{int}c_\mathrm{int}L_\mathrm{int}^2,
\end{split}
\end{equation}
where $R_\mathrm{S}$ and $C_\mathrm{S}$ are the source resistance and capacitance, respectively, $C_\mathrm{L}$ is the load resistance (assumed equal to $C_\mathrm{S}$). The energy dissipation of the CMOS circuit is given as
\begin{eqnarray}
E_\mathrm{CMOS} = \left(C_\mathrm{S}+C_\mathrm{L}+c_\mathrm{int}L_\mathrm{int}\right)V_\mathrm{DD}^2,
\end{eqnarray} 
where $V_\mathrm{DD}$ is the supply voltage. 

CMOS device metrics are taken from the ITRS roadmap for the 2020 technology node (1/2 pitch of metal-1 = 18 nm). For a minimum-sized inverter, $R_\mathrm{S} \approx$ 78 k$\Omega$, $C_\mathrm{S}$ = 0.68 fF/$\mu$m, $C_\mathrm{L}$ = 0.38 fF/$\mu$m, $\rho_\mathrm{eff}$ = 25 $\mu\Omega$cm, $c_\mathrm{int}$ = 1.6 pF/cm, interconnect aspect ratio = 2~\cite{wilson2013international}. The delay of the CMOS circuit by omitting interconnect related delay is $\approx$ 1.1 ps at an energy dissipation of 10 aJ/bit. This yields an EDP of 1.1$\times 10^{-29}$ Js. 
For an interconnect length of 100 nm, the delay of the CMOS circuit is 2.1 ps at an energy dissipation of 20 aJ/bit and EDP of 4.2$\times 10^{-29}$ Js. 

Existing spin-based devices that are considered for comparison include all-spin logic (ASL)~\cite{behin2010proposal}, charge-spin logic (CSL)~\cite{datta2012non}, magnetoelectric spin-orbit (MESO) logic~\cite{manipatruni2015spin}. ASL uses filtering of electric current through a nanomagnet to generate spin-polarized current, which communicates spin information between input-output nanomagnets via a non-magnetic conductor (e.g. copper, aluminum) that serves as the interconnect.
Unlike charge current, spin current is not conserved; therefore, the design of interconnects in ASL requires careful consideration~\cite{rakheja2013impact}. 
On the other hand, CSL uses spin-Hall effect to convert electric current carrying information into spin polarized current, which is used to switch the state of an input nanomagnet. 
The orientation of the input nanomagnet is communicated to an output nanomagnet via their mutual magnetic dipolar coupling.
The magnetization state of the output nanomagnet is read through an MTJ, which generates an output electric current with the polarity and amplitude dependent on the orientation of the output nanomagnet and the voltage applied on the MTJ.  
Since information is communicated via electric current, there is no loss of information in the interconnect. 
However, due to the flow of electrical current through a heavy metal layer with a high effective resistivity, CSL has a high Joule heating. 
The MESO device, recently proposed in~\cite{manipatruni2015spin}, uses magnetoelectric transduction to convert electrical current into spin current at the input side, while spin-orbit coupling is utilized at the output end for spin to charge transduction. That is, the input and output state variables are encoded in electrical current. Benchmarking activities have shown that magnetoelectric mediated spin devices have energy dissipation comparable to that of CMOS~\cite{nikonov2015benchmarking}.

Table~\ref{tab:metric} shows the performance metrics of vTOPSS in comparison to spin-based devices. The performance of vTOPSS exceeds that of existing spin-based devices.  
The energy dissipation of vTOPSS is 100$\times$ lower than that of ASL and CSL, while the delay of vTOPSS is (2-10)$\times$ lower than that of ASL and CSL, respectively.
The delay of vTOPSS is comparable to that of the MESO device and can be reduced further by utilizing MI layers with lower damping and/or MI switching via precessional dynamics. 
In terms of energy dissipation, vTOPSS performs slightly better than the MESO device. The energy dissipation can be further reduced through material optimization, particularly with a higher TMR and RA product of the MTJ used for sensing the state of the MI  layer in vTOPSS.

\begin{table*}
\begin{center}
\caption{Overview of performance metrics of various spin-based devices. Performance metrics of vTOPSS exceed those of existing spin-based technologies. ASL: All-spin logic, CSL: Charge-spin logic, MESO: Magnetoelectric spin orbit logic. The EDP of low-power CMOS technology at the 2020 ITRS technology node is $\approx 4\times 10^{-29}$ Js (see text for calculations). $^*$indicates results for perpendicular magnetic anisotropy magnets, $^{**}$total pulse width reported in~\cite{manipatruni2015spin}, $^{\#}$conservative estimate as it assumes areas of TI layer and the MTJ are equal to (100$\times$100) nm$^2$ and spacing between TI and MTJ = 50 nm. vTOPSS area is calculated for the device in Fig.~\ref{fig:device}.}
\renewcommand*{\arraystretch}{1.3}
\begin{tabular}{p{2.5cm}|p{3.5cm}|p{3.5cm}|p{3.5cm}|p{3.5cm}}
\hline
\bf{Metric} & \bf{ASL~\cite{manipatruni2016material}} & \bf{CSL~\cite{rangarajan2017energy}} & \bf{MESO~\cite{manipatruni2015spin}} & \bf{vTOPSS (this work)} \\
\hline
Input/Output & Voltage & Electrical current & Electrical current & Voltage \\
Transduction & $V\rightarrow$ \bf{m} $\rightarrow I_\mathrm{spin} \rightarrow$ \bf{m} $\rightarrow V$ & $I_\mathrm{elec} \rightarrow$ \bf{m} $\rightarrow I_\mathrm{elec}$ & $I_\mathrm{elec} \rightarrow$ \bf{m} $\rightarrow I_\mathrm{elec}$ & $V \rightarrow$ \bf{m} $\rightarrow V$ \\
Energy-per-bit & 0.34 fJ(*) & 0.32 fJ & 27 aJ & 21 aJ \\
Switching delay & 0.5 ns & 1.5 ns & 250 ps(**) & 200 ps \\
Energy-delay product & 17$\times 10^{-26}$ Js & 4.8$\times 10^{-25}$ Js & 6.75$\times 10^{-27}$ Js & 4.2$\times 10^{-27}$ Js \\
Area & 3.8$\times 10^{-3}$ $\mu$m$^2$ & 1.6 $\times 10^{-3}$ $\mu$m$^2$ & 1.4 $\times 10^{-2}$ $\mu$m$^2$ & (1-2) $\times 10^{-2}$ $\mu$m$^2$(\#) \\
Fan-out & No & Yes & Yes & Yes \\
\hline
\multicolumn{5}{p{16.5cm}}{vTOPSS material parameters: $\mathcal{A}_\mathrm{int}$ = $\mathcal{A}_\mathrm{MTJ}$ = (100$\times$100) nm$^2$, $\alpha$ = 3$\times 10^{-3}$, $\sigma_\mathrm{SHC}$ = 2000$\hbar/2e$ $\Omega^{-1}$cm$^{-1}$, $E_b$ = 30$kT$, $H_K$ = 0.1 T, $\varepsilon$ = 0.5, $c_\mathrm{int}$ = 1.6 pF/cm, $\rho_{eff}$ = 25 $\mu\Omega$cm, TMR = 1.4, RA = 100 $\Omega\mu$m$^2$, interconnect length and width = 100 nm, $G_\mathrm{TI}$ = 0 (no leakage), $V_\mathrm{read}$ = 50 mV.} \\
\hline
\end{tabular}
\label{tab:metric}
\end{center}
\end{table*}

\vspace{-10pt}   
\section{Universal Boolean Logic Implementation}
\begin{figure}[h!]
\centering
\includegraphics[width=3.25in]{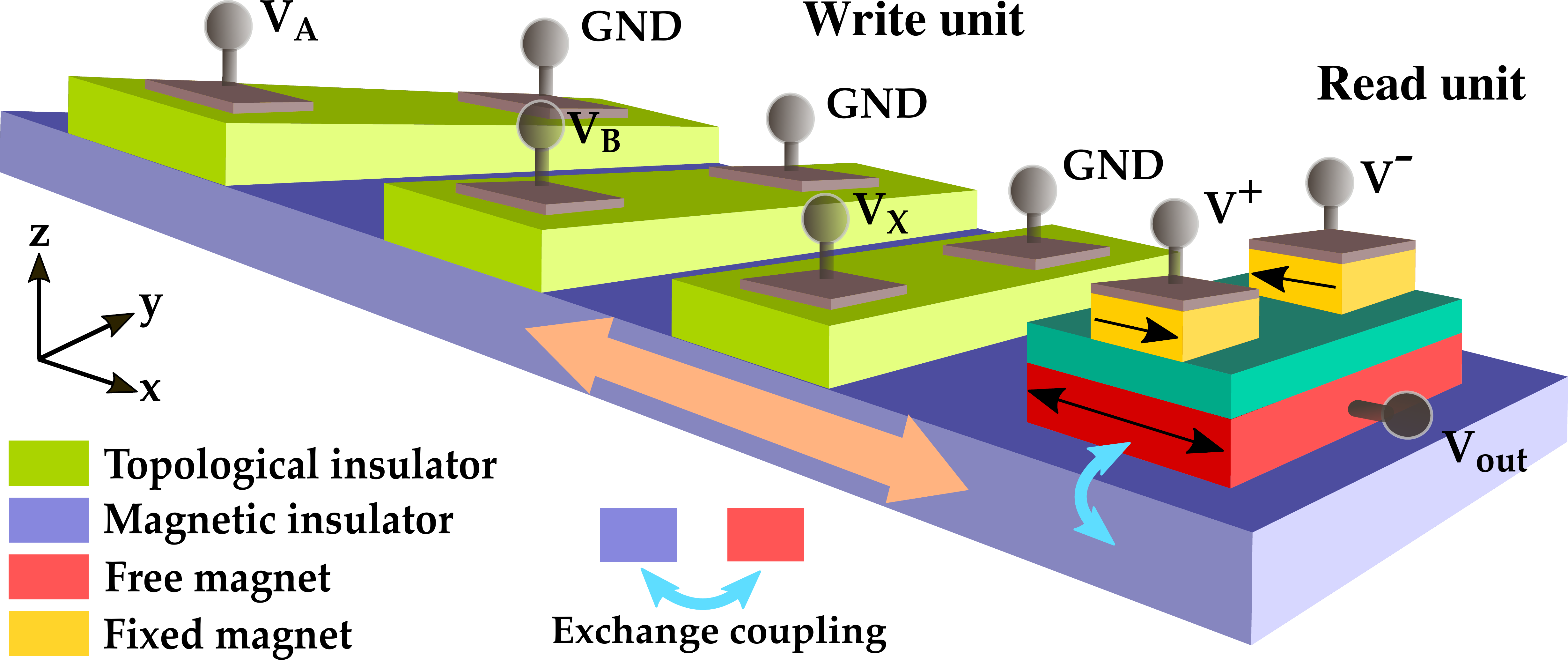} 
\caption{All two-input Boolean logic functions can be implemented using the same device layout. The primary inputs are denoted as $V_\mathrm{A}$ and $V_\mathrm{B}$, while the signal $V_\mathrm{X}$ denotes the tie-breaking signal to change the Boolean functionality. To switch between inverting and non-inverting logic (different polarities of $V_\mathrm{out}$), the polarity of the signals $V^+$ and $V^-$ at the MTJ can be interchanged.}
\label{fig:nand}
\end{figure}

\vspace{-10pt}
A complete set of two-input Boolean functions can be implemented using the schematic shown in Fig.~\ref{fig:nand}. In this figure, $V_\mathrm{A}$ and $V_\mathrm{B}$ refer to primary signal inputs, while $V_\mathrm{X}$ denotes the tie-breaking input signal. To change the functionality between true and complementary outputs, the polarity of the supply voltage signals on the MTJ is swapped. To implement NAND gate, $V_\mathrm{X}$ is set to its negative value, while for NOR gate, $V_\mathrm{X}$ is set to its positive voltage. For XOR/XNOR functionality, one of the primary inputs is applied as a voltage signal on the TI, while the other primary input will serve as the supply voltage on the MTJ in the read unit. In the case of copy/invert functions, the tie-breaking signal $V_\mathrm{X}$ is set to 0 V. Alternately, the schematic shown in Fig.~\ref{fig:device} can be used for copy and invert Boolean operations. However, by using a generic layout as in Fig.~\ref{fig:nand}, all 16 Boolean operations possible for two input signals can be implemented directly by permuting the polarities of MTJ supply and the control voltage. 
Another major advantage of vTOPSS is its ability to support 
%Due to its innate polymorphism, vTOPSS supports 
logic locking~\cite{xie2016mitigating} and encryption at the device level by preventing optical based reverse engineering attacks~\cite{wang2017probing}. The innate polymorphism of vTOPSS will enable runtime reconfigurability where the actual function to be implemented is determined on-the-fly using a {\textit{key}}/control input. Exploring the resilience of vTOPSS against existing hardware attacks, prominently those based on the Boolean satisfiability test, will be investigated in future work.

\begin{figure}[h!]
\centering
\includegraphics[width=3.5in]{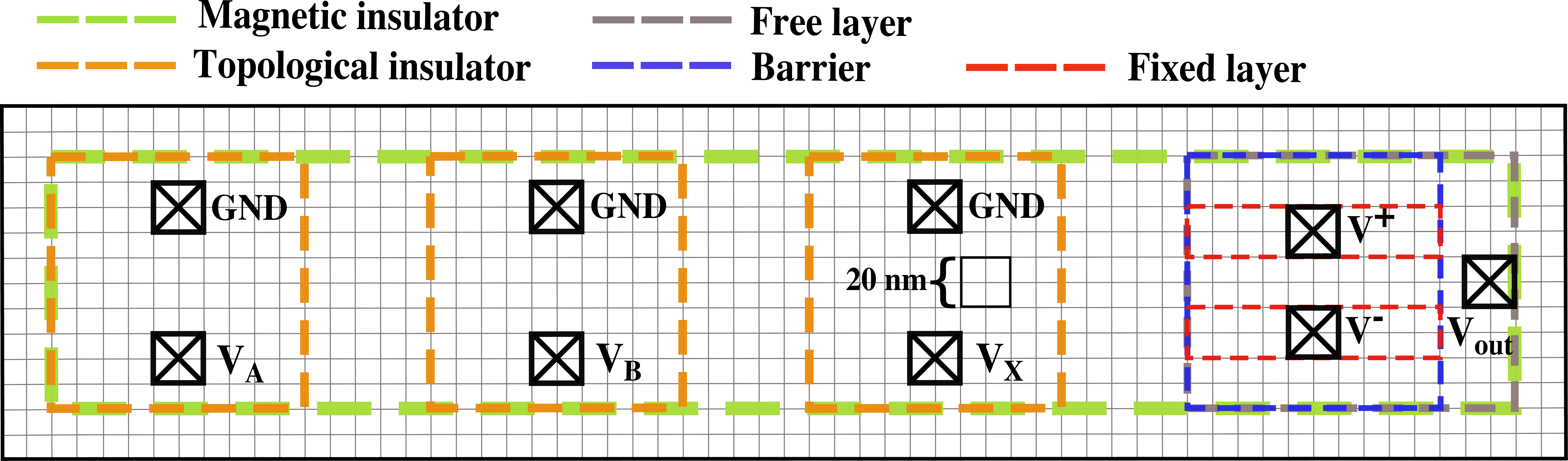} 
\caption{Device layout for schematic shown in Fig.~\ref{fig:nand}. Here, it is assumed that the cross-sectional area of the TI layer is (100$\times$100) nm$^2$ and the spacing between adjacent TI layers is 50 nm. The MTJ cross-sectional area is the same as that of the TI-MI interface. The total area is $\approx$ 0.06 $\mu$m$^2$. }
\label{fig:layout}
\vspace{-10pt}
\end{figure}

The device layout corresponding to the universal logic gate is shown in Fig.~\ref{fig:layout}, where the device area for a universal gate is $\approx$ 0.06 $\mu$m$^2$ assuming relatively large values of the cross-sectional areas of the TI-MI interface, MTJ, and the interconnect. The area can be reduced significantly by patterning narrower TI/MI layers and reducing the cross-sectional dimensions of the MTJ. The latter approach, in particular, is advantageous for reducing the device footprint without a negative impact on the device performance metrics.  

\vspace{-10pt}
\section{Conclusions}
\vspace{-10pt}
Computational electronics can, in principle, be realized using any state variable that is stable over device-relevant timescales, and with any low-loss communication mechanism between devices that allows fan-out. 
In this regard, storing and manipulating information in magnetic materials is promising. Magnetic materials have a large number of electron spins that are locked together by their exchange interaction such that the reorientation energy per spin to move the magnetization collectively can be on the order of meV. 
Magnetization reversal solely through electric fields is critical toward paving the path for an ultra-low-energy computing substrate. 
The efficiency of voltage-spin conversion must be significantly higher to allow ultra-low-voltage operation to be competitive with CMOS technology.

In this paper, 
a voltage-controlled topological spin switch (vTOPSS) based on a hybrid toplogical insulator (TI)-magnetic insulator (MI) magnetoelectric structure was presented. The device has the following important features: (i) innate polymorphism, i.e. it can implement all 16 two-input Boolean operations using the same layout, (ii) CMOS compatible (input/output variables are in voltage domain), (iii) extremely large intrinsic gain for charge-to-spin conversion owing to the ultra-high spin Hall conductivity of the TI material, (iv) ability to support fan-out, (v) sub-10 mV operation with energy-per-bit $<$ 10 aJ/bit, (vi) ability to lower EDP on the order of $10^{-29}$ Js (competitive with CMOS), (vii) elimination of electrical current carrying wires as the operation is fully  based on voltage-to-voltage conversion with transmission of information via capacitive charging/discharging of wires, (viii) ultra-low damping of the MI layer allows ultra-fast operation on the order of few 100's of picoseconds via anti-damping spin torque.

We developed analytic models to quantify the performance of vTOPSS and benchmark the results against existing CMOS and spin-based devices. Our results conclusively show that at the current state-of-the-art material parameters, vTOPSS exceeds the performance of all-spin logic, charge-spin logic, and magnetoelectric spin-orbit logic. Improvements in material parameters and device design can readily facilitate sub-aJ energy-per-bit operation with an energy-delay product $\sim$ $10^{-29}$ Js for vTOPSS to be competitive against CMOS devices at the 2020 ITRS technology node. Future work will address important issues pertinent to multi-domain effects in both uniaxial and biaxial magnetic insulators and effects of thermal stochasticity for sub-critical excitation. 
Unlike CMOS, vTOPSS can also provide logic locking due to the uniform device-level layout that makes it virtually impossible to probe the functionality with reverse engineering hardware attacks. The ability of vTOPSS to thwart state-of-the-art Boolean SAT attacks is yet to be examined and will be considered in future work.

\vspace{-10pt}
\begin{acknowledgments}
\vspace{-10pt}
This work was supported partially by the MRSEC Program of the National Science Foundation under Award Number DMR-1420073.
A. D. Kent acknowledges support from NSF-DMR-1610416. 
The authors would like to thank Prof. Nitin Samarth at Pennsylvania State University for useful discussions during the preparation of the manuscript. S. Rakheja would also like to thank Nikhil Rangarajan at New York University for his help in generating some of the graphics in the paper. 
\end{acknowledgments}

%\bibliography{apssamp,DevicePhysics-mef}
\providecommand{\noopsort}[1]{}\providecommand{\singleletter}[1]{#1}%

\end{document}